\documentclass[prb,twocolumn,aps,showpacs]{revtex4}
\usepackage{graphicx}
\usepackage{bm}
\usepackage{amssymb} 
\begin{document}

\title{Weakly chiral networks and 2D delocalized states in a weak magnetic field}

\author{V. V. Mkhitaryan$^{1}$, V. Kagalovsky$^{2}$, and M. E. Raikh$^{1}$}

\affiliation{$^{1}$Department of Physics, University of Utah, Salt
Lake
City, UT 84112, USA\\
$^{2}$Sami Shamoon College of Engineering, Beer Sheva 84100,
Israel}

\begin{abstract}
We study numerically the localization properties of
two-dimensional electrons in a weak perpendicular magnetic field.
For this purpose we construct {\em weakly chiral} network models
on the square and triangular lattices. The prime idea is to
separate in space the regions with {\em phase} action of magnetic
field, where it affects  interference in course of multiple
disorder scattering, and the regions with {\em orbital} action of
magnetic field, where it bends electron trajectories. In our
models, the disorder mixes counter-propagating channels on the
links, while scattering matrices at the nodes describe exclusively
the bending of electron trajectories. By artificially introducing
a strong spread in the scattering strengths on the links (but
keeping the average strength constant), we eliminate the
interference and reduce the electron propagation over a network to
a classical percolation problem. In this limit we establish the
form of the \hspace{2mm}  {\it disorder -- magnetic field} \hspace{2mm} phase diagram. This
diagram contains the regions with and without edge states, i.e.
the regions with zero and quantized Hall conductivities. Taking
into account that, for a given disorder, the scattering strength
scales as inverse electron energy, we find agreement of our phase
diagram with levitation scenario: energy separating the Anderson
and quantum Hall insulating phases floats up to infinity upon
decreasing magnetic field. From numerical study, based on the
analysis of quantum transmission of the network with random phases
on the links, we conclude that the positions of the weak-field
quantum Hall transitions on the phase diagram are very close to
our classical-percolation results. We checked that, in accord with
the Pruisken theory, presence or absence of time reversal symmetry
{\it on the links} has no effect on the line of delocalization
transitions.  We also find that floating up of delocalized states
in energy is accompanied by {\em doubling} of the critical
exponent of the localization radius. We establish the origin of
this doubling within classical-percolation analysis.
\end{abstract}
\pacs{72.15.Rn; 73.20.Fz; 73.43.-f}
\maketitle

\section{Introduction}

\subsection{Levitation scenario}

Scaling theory of localization \cite{AALR} predicts that evolution
with size, $L$, of the conductivity, $\sigma$, (in the units of
$e^2/2\pi\hbar$) of a 2D sample is governed only by the value of
$\sigma$, regardless of the type of disorder in the sample, i.e.,
\begin{equation}
\label{scaling} \frac{\partial\sigma}{\partial\ln
L}=\sigma\cdot\beta(\sigma).
\end{equation}
Together with initial condition, $\sigma{\big |}_{L\sim
l}=\sigma_0= k_{\scriptscriptstyle F}l$, where
$k_{\scriptscriptstyle F}$ is the Fermi momentum, and $l$ is the
transport mean free path, Eq. (\ref{scaling}) suggests that, in
zero magnetic field, where $\beta(\sigma)=-2/(\pi\sigma)$,
localization radius of electron states is given by
\begin{equation}
\label{orthogonal}
\ln\Bigl(\frac{\xi_o}{l}\Bigr)=\frac{\pi\sigma_0}2=\frac\pi2
k_{\scriptscriptstyle F}l.
\end{equation}
With increasing magnetic field, when $\beta(\sigma)$ crosses from
the orthogonal to the unitary form,
$\beta_u(\sigma)=-1/(\pi\sigma)^2$, the localization radius
crosses over from $\xi_o$ to $\xi_u$, given by
\begin{equation}
\label{unitary}
\ln\Bigl(\frac{\xi_u}{l}\Bigr)=\pi^2\sigma_0^2=\pi^2(k_{\scriptscriptstyle
F}l)^2.
\end{equation}
Crossover takes place when weak localization is suppressed, i.e.,
when the magnetic flux through the vector area spanned by the
electron travelling diffusively over the area $\sim \xi_o^2$, is
of the order of the flux quantum. With vector area being
$\sim\xi_ol$, we get the following estimate for crossover magnetic
field
\begin{equation}
\label{estimate} \omega_c\tau=\frac{l}{\xi_o} \sim
\exp\left(-\frac\pi2 k_{\scriptscriptstyle F}l\right),
\end{equation}
where $\omega_c$ is the cyclotron frequency, and
$\tau$ is the scattering time.
\begin{figure}[t]
\centerline{\includegraphics[width=85mm,angle=0,clip]{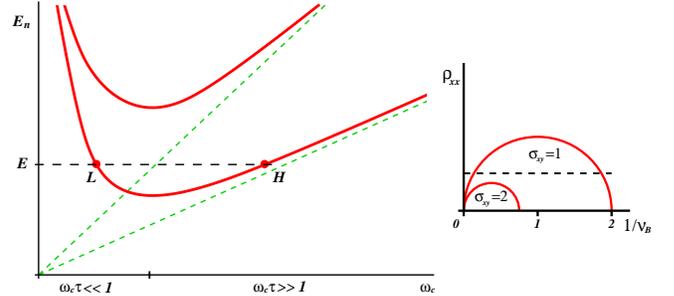}}
\caption{(Color online) Left: Energy position of two lowest
delocalized states, $E_0$ and $E_1$, as a function of magnetic
field, $\omega_c$, as predicted in Ref.~\onlinecite{Khmelnitskii}.
At a given $E$, delocalization transition occurs at low field
(point $L$) and at high field (point $H$). Right: Same
dependencies, replotted in the axis inverse filling factor vs.
zero-field resistivity, constitute a part of the global phase
diagram Ref. \onlinecite{Kivelson}. Depending on $\rho_{xx}^0$,
the system undergoes a sequence of transitions,
$0\rightarrow1\rightarrow0$ or
$0\rightarrow1\rightarrow2\rightarrow1\rightarrow0$.}
\label{levit}
\end{figure}

The origin of the crossover Eq.~(\ref{estimate}) is that the
paths, which interfere in a zero field, acquire field-induced
random Aharonov-Bohm phases.
Justification for considering exclusively the {\it phase} action
of magnetic field is that in high-mobility samples with
$k_{\scriptscriptstyle F}l\gg 1$ the crossover field is so weak
that its orbital action can be neglected. It is a very delicate
fact that, after the crossover, this orbital action causes a
drastic change of the eigenstates even for classically weak
magnetic fields, $\omega_c\tau \ll 1$. This conclusion was drawn
by Khmelinitskii \cite{Khmelnitskii} from the analysis of the
renormalization group flows \cite{Khmelnitskii, pru83}.
Originally, these equations were derived to describe quantization
of the Hall conductivity in a strong-field limit, $\omega_c\tau
\gg 1$,
\begin{eqnarray}
\label{pru1}
&&\frac{\partial\sigma_{xx}}{\partial\ln L}=-\frac{1}{2\pi^2\sigma_{xx}}-
\sigma_{xx}^2{\cal D}e^{-2\pi\sigma_{xx}}\cos(2\pi\sigma_{xy}),\\
&&\frac{\partial\sigma_{xy}}{\partial\ln L}=
-\sigma_{xx}^2{\cal D}e^{-2\pi\sigma_{xx}}\sin(2\pi\sigma_{xy}),\label{pru2}
\end{eqnarray}
where $\sigma_{xx}$ and $\sigma_{xy}$ are, respectively, the
diagonal and non-diagonal components of the conductivity tensor,
and ${\cal D}$ is a dimensionless constant. First term of Eq.
(\ref{pru1}) is the same as in Eq. (\ref{scaling}) with unitary
$\beta(\sigma)$. It originates from interference: two paths
corresponding to the {\em same} scatterers but different sequences
of scattering events interfere even in the presence of
Aharonov-Bohm phases. Second term reflects the orbital action of
magnetic field (Lorentz force); by curving electron trajectories
it tends to destroy the interference. Quantum Hall transition
between $\sigma_{xy}{\big |}_{L\rightarrow \infty}=n$ and
$\sigma_{xy}{\big |}_{L\rightarrow \infty}=n+1$ takes place when
the ``phase'' and ``orbital'' terms compensate each other.
Khmelnitskii's  treatment \cite{Khmelnitskii} is equivalent to
solving Eqs. (\ref{pru1}), (\ref{pru2}) together with classical
Drude initial condition,
\begin{equation}
\label{Drude} \sigma_{xx}{\big |}_{L\sim
l}=\frac{\sigma_0}{1+(\omega_c\tau)^2},\qquad \sigma_{xy}{\big
|}_{L\sim l}=\frac{\sigma_0\,(\omega_c\tau)}{1+(\omega_c\tau)^2},
\end{equation}
which yields the positions of delocalized states
\begin{equation}
\label{positions} E_n =\hbar\omega_c\left(n+\frac{1}{2}\right)
\left[1+\frac1{(\omega_c\tau)^{2}}\right].
\end{equation}
As shown in Fig.~\ref{levit} for $n=0,\,1$, the high-field part,
$\omega_c\tau\gg 1$, of $E_n$ follows the centers of
Landau levels, while the low-field part,
$E_n\tau\approx(n+1/2)(\omega_c\tau)^{-1}$, ``floats up'' as
$\omega_c\tau \rightarrow 0$.
Such a behavior of critical values of $k_{\scriptscriptstyle
F}l=E_{\scriptscriptstyle F}\tau$ in vanishing field is usually
called levitation of delocalized states \cite{Lev_1}. More
specifically, it is expected that, upon increasing magnetic field
above the crossover Eq. (\ref{estimate}), localization radius,
$\xi(\omega_c)$, diverges in the vicinity of discrete values
$\omega_c\tau=(n+1/2)(k_{\scriptscriptstyle F}l)^{-1}$, changing
from unitary $\xi_u$ to infinity and returning back to $\xi_u$.
Recasting Eq. (\ref{positions}) into the dependence,
$\rho^0_{xx}=\sigma_0^{-1}$, versus the inverse filling factor,
$\nu^{-1}_{\scriptscriptstyle B}=\omega_c/(2E_{\scriptscriptstyle
F})$, yields a system of semicircles
\begin{equation}
\rho_{xx}^2+\left(\frac1{\nu_{\scriptscriptstyle B}}
-\frac1{n+1/2}\right)^2=\frac1{(n+1/2)^2},
\end{equation}
shown in Fig. \ref{levit} inset, which is a part of the global
phase diagram \cite{Kivelson}. This diagram suggests that for high
enough $\rho^0_{xx}$, i.e., for strong disorder, the resistance
$\rho_{xx}(\omega_c)$ grows monotonically. Upon decreasing
$\rho^0_{xx}$, below a certain threshold value, Fig.
 \ref{levit}
(e.g., by applying the gate voltage) $\rho_{xx}(\omega_c)$
exhibits two quantum Hall peaks.  For $\rho^0_{xx}$ smaller than
the second threshold, Fig. \ref{levit}, $\rho_{xx}(\omega_c)$
exhibits four peaks, and so on. By now, such a behavior (two peaks
in $\rho_{xx}(\omega_c)$ dependence for small enough
$\rho^0_{xx}$) was reported in a number of experimental papers
Refs. \onlinecite{Jiang92, Jiang93, Jiang93PRL, Jiang95PRL,
Jiang95, Jiang98, Kravchenko95, Wang94, Hughes94, ShaharLev}. On
the other hand, theoretical numerical studies \cite{xiong01,
liu96, yang96, sheng97, sheng98, sheng00, koschny01,Wen,wan01,Sch}
aimed at revealing levitation on a microscopic level, are less
conclusive \cite{Huckestein00}. The only established fact is the
{\em tendency} \cite{Shahbaz95, Kagalovsky95, kagalovsky97,
Gramada96, Haldane97, Fogler98,Schulz} for floating up of
delocalized states upon decreasing $\omega_c$ in the {\em
strong-field domain}, where Landau levels are still well-defined.
This tendency is due to disorder-induced mixing of the neighboring
well-resolved Landau levels.

\subsection{Network models}

First numerical verification \cite{KramMac} of one-parameter
scaling Eq. (\ref{scaling}) was performed within the Anderson
model \cite{Anderson58}. Physically, this model  corresponds to
realizations of disorder in which scatterers have random strength,
while positional disorder is eliminated by placing scatterers on
the lattice. In other words, the phase acquired by electron
between two subsequent scattering acts is assumed to be the same.
Important is that the minimal model \cite{Anderson58}, in which
disorder is characterized by a single dimensionless parameter,
spread of the site energies in the units of bandwidth,  captures
all features of a generic random potential.

\begin{figure}[t]
\centerline{\includegraphics[width=60mm,angle=0,clip]{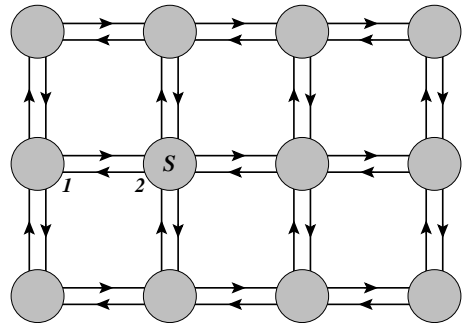}}
\caption{In non-chiral network model, confining potential
restricts electron motion to the links. All the scattering
matrices $S$ at the nodes are the same. Positional disorder is
emulated by randomness of phases acquired on the links. Depending
on symmetry class, the phases between the points 1 and 2 are
either the same or different. } \label{network}
\end{figure}

Another minimal description of disorder is at the core of
scattering approach to localization introduced by B.~Shapiro
\cite{Shapiro}. Microscopic realization of  the network
Ref.~\onlinecite{Shapiro} requires restricting  the electron
motion by a  confining potential, as illustrated in Fig.
\ref{network}. This confining potential ensures that there are
only four possible outcomes of scattering, which takes place at
the nodes. In contrast to Anderson model, all scatterers at the
nodes of network are assumed identical, while positional disorder
is maximally strong. This is achieved by assuming that phases,
accumulated between the neighboring nodes, are completely random.
Within the network model description,
a physical parameter, $(k_{F}l)^{-1}$, is emulated by
$1-T$, where $T$ is the transmission of the node. One of the
apparent successes of the network model description of disordered
systems was the demonstration \cite{SO} of the
zero-field localization-delocalization transition in 2D system with
spin-orbit scattering. Formal origin of this transition is the change
of sign of $\beta(g)$ in Eq. (\ref{scaling})
in the presence of  spin-orbit scattering \cite{HikLar}

Network-model approach is especially well suited for the
description of the quantum Hall transition in a strong
magnetic field. As was pointed out by Chalker and Coddington
\cite{CC} (CC), in this case, unlike zero magnetic field, the
links of the network acquire a natural physical meaning, namely,
they coincide with  equipotential lines \cite{Kazarinov,
Iordansky, Trugman} of the {\it bare} smooth (on the scale of
magnetic length) random potential. This is due to the
field-induced quenching of kinetic energy of electron, rather than
due to artificially imposed confining potential. In addition, in
strong magnetic field, the motion along each link, representing
the drift of the Larmour circle, is {\it unidirectional}. Nodes in
the of the CC network also have a transparent meaning: they
represent saddle points of the random potential, where
equipotentials come as close as magnetic length. Various aspects
of the Quantum Hall transition, relevant to experiment
\cite{wei90,pan97,schaijk00,koch91,
Tsui05,Tsui09,Cobden96,hohls02,peled02}, {\it e.g.}, divergence of
the localization radius (scaling \cite{CC,Kivelson93,Ohtsuki09}),
critical statistics of energy levels \cite{klesse},  mesoscopic
conductance fluctuations \cite{Jovanovic96, Fisher97},
point-contact conductance \cite{zirnbauer99}, were studied
theoretically using the CC model \cite{KramOhtsKet}.

Testing the levitation scenario microscopically requires to
construct a minimal {\em weakly chiral} network model, which
captures the physics encoded in the system Eq.~(\ref{pru1}),
namely competition between interference-induced localization and
orbital-induced curving asymmetry in the scattering to the
``left'' and to the ``right''. Construction of such a network and
study of its localization properties is the objective of the
present paper. On the physical grounds, the desired description
should contain only two parameters, $k_{\scriptscriptstyle F}l$
and $\omega_c\tau$. Short communication on the results reported
below can be found in Ref. \onlinecite{We}.

\section{Reformulation of non-chiral network model}

We achieve the goal of constructing a minimal weakly-chiral
network model in two steps. First we reformulate the standard
non-chiral network model Fig. \ref{network} by separating each
node into the regions with backscattering and left-right
scattering. As a second step we incorporate weak chirality in the
form of imbalance between scattering to the left and scattering to
the right.

In a standard non-chiral network model Fig. \ref{network} the
scattering at the node is described by  $4\times 4$ unitary
scattering matrix. This matrix  can be parameterized by three
independent numbers, {\it e.g.}, as follows:
\begin{eqnarray}
\label{gsm}
S=\left(\begin{array}{cccc} r&-d_1&-t&-d_2\\
d_1&r&d_2&-t\\t&-d_2&r&d_1\\d_2&t&-d_1&r
\end{array}\right),
\end{eqnarray}
where reflection, $r$, transmission, $t$, and deflection
coefficients, $d_1$, $d_2$, are all real and satisfy the flux
conservation condition
\begin{equation}
r^2+t^2+d_1^2+d_2^2=1.
\end{equation}
Isotropy requires that $d=d_1=d_2$, so that the scatterer is
characterized by only two independent parameters, say $t$ and $d$.
The model belongs to the orthogonal symmetry class if the phase,
$\varphi_{12}$, accumulated upon propagation $1\rightarrow 2$
between the scatterers $1$ and $2$, Fig. \ref{network}, is equal
to the phase, $\varphi_{21}$, accumulated upon propagation
$2\rightarrow 1$. Otherwise, it belongs to the unitary class.
\begin{figure}[t]
\centerline{\includegraphics[width=90mm,angle=0,clip]{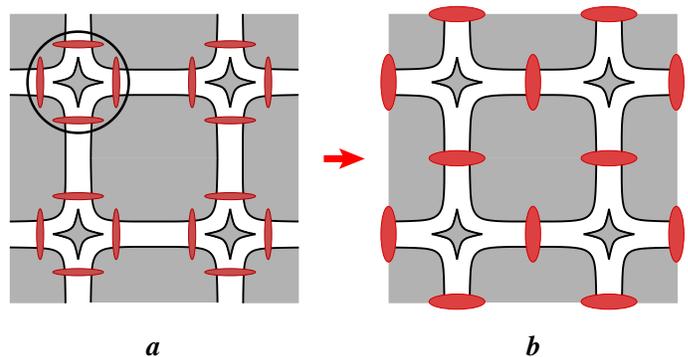}}
\caption{(Color online) Construction of weakly chiral  network
model. a: Node scattering matrix is decomposed into four
"reflectors" and a "junction". b: Final network model upon
combining two reflectors on a given link into one reflector.}
\label{reform}
\end{figure}
\begin{figure}[b]
\centerline{\includegraphics[width=85mm,angle=0,clip]{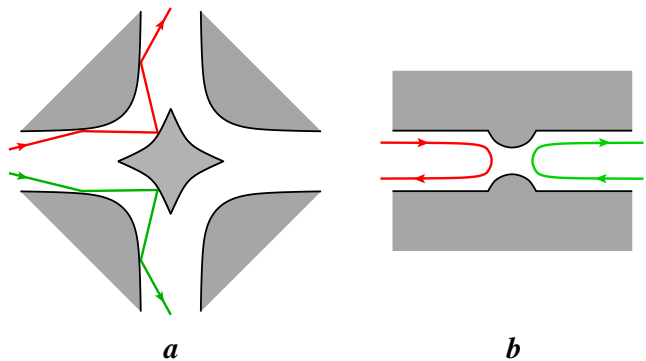}}
\caption{(Color online) a: Microscopic realization of the junction
matrix Eq.~(\ref{S0}). The junction is defined by the confinement
potential (shaded area); rhomboidal scatterer at the center
ensures that each incident wave is deflected {\it only} to the
left or to the right. b: Scattering on a link, described by matrix
Eq.~(\ref{Matrixp}), can be modelled with a point contact.}
\label{junction}
\end{figure}

Full localization of the eigenstates in 2D, predicted by the
scaling theory \cite{AALR}, manifests itself in vanishing
transmission of the network for {\it all} sets of $t$ and $d$. The
degree of localization is governed by the conductance \cite{SO}
\begin{equation}\label{mfp}
k_{\scriptscriptstyle F}l=\frac12\,\frac{t^2+d^2}{1-t^2-d^2}.
\end{equation}
If $d$ and $t$ are small, localization is strong (within one
plaquette). Conversely, for $t$ close to $1$, electron changes the
direction of propagation after $\sim(1-t^2)^{-1}\gg1$ scattering
acts, which corresponds to high conductance, $
k_{\scriptscriptstyle F}l\gg1$, and exponentially large
localization radius. It is important to mention one particular
case, namely, $t\ll1$ and $d\approx 1/\sqrt{2}$. According to Eq.
(\ref{mfp}), this case of almost complete deflection, with weak
reflection and transmission, should correspond to strong
localization. This is, however, not the case. In fact, the states
become progressively delocalized as $d$ approaches the value
$1/\sqrt{2}$. This conclusion can be drawn from Ref.
\onlinecite{Bocquet}, where the corresponding limit of the network
model has been studied. Since the case $d\approx 1/\sqrt{2}$ will
play an important role in our construction later on, we discuss a
seeming contradiction to the scaling in this case in Appendix.

Note in passing, that the parametrization Eq. (\ref{gsm}) reflects
the Born scattering (for $d_1=d_2$), where the probabilities of
deflection, forward, and backward scattering are independent of
the direction of incidence. This is certainly not the general case
of potential scattering. In fact, one can relax the condition
$d_1=d_2$ for a given node and ensure global isotropy by requiring
that the scattering to the left and to the right are equally
probable {\it on average} (over the nodes).

Our reformulation of the network model is illustrated in Fig.
\ref{reform}. We start with  specifying  scattering matrices at
the nodes of Fig. \ref{network} as systems of four "reflectors"
and one junction, Fig. \ref{reform}. Each reflector mixes only two
channels on the corresponding link, so that its scattering matrix
\begin{equation} \label{Matrixp0} {\cal P}_0=
\left(\begin{array}{cc}
\sqrt{1-p_0}&\sqrt{p_0}\\
\,\\
 -\sqrt{p_0}&\sqrt{1-p_0}
\end{array}\right)
\end{equation}
is $2\times2$, where $p_0$ is the power reflection coefficient.
The junction, Fig. \ref{junction}, {\it does not} transmit or
reflect incoming waves, but rather scatters them either to the
left or to the right with equal probability, $1/2$. The
corresponding $4\times 4$ scattering matrix has the form
\begin{equation}
\label{S0} {\cal S}_0= \left(\begin{array}{cccc}
0 &-\frac1{\sqrt{2}}&0 &-\frac1{\sqrt{2}}\\
\frac1{\sqrt{2}}&0 &\frac1{\sqrt{2}}&0\\
0 &-\frac1{\sqrt{2}}&0 &\frac1{\sqrt{2}}\\
\frac1{\sqrt{2}}&0 &-\frac1{\sqrt{2}}&0\\
\end{array}\right).
\end{equation}
It is important to demonstrate that the combination of a junction
and four reflectors which is characterized by a single parameter,
$p_0$, corresponds to the effective node of the scattering matrix
with appropriate symmetry.
Let us denote with $\phi_i$ the phase accumulated between the
junction and reflector on the link $i$, $i=1,..,4$, Fig.
\ref{reform}a. For potential scattering with time-reversal
symmetry preserved, the phase accumulated between the reflector
and the junction is the same $\phi_i$, as for propagation between
the junction and reflector. Then the effective node scattering
matrix assumes the form
\begin{eqnarray}
\label{arbphase}
{\cal S}=\left(\begin{array}{cccc} R^1_{2,4}&-D_{1,2}&T_{4,2}&D_{1,4}\\
\,\\
D_{2,1}&R^2_{3,1}&D_{2,3}&T_{3,1}\\
\,\\
T_{2,4}&-D_{3,2}&R^3_{4,2}&D_{3,4}\\
\,\\
D_{4,1}&T_{1,3}&-D_{4,3}&R^4_{1,3}
\end{array}\right),
\end{eqnarray}
where the deflection, transmission, and reflection coefficients
are defined as
\begin{eqnarray} \label{def1} &&D_{i,j}=
\frac{\sqrt{2}(1-p_0)}{\cal N}\bigl(e^{i(\phi_i+\phi_j)}+p_0e^{i(2\psi-\phi_i-\phi_j)}\bigr),\\
&&\,\nonumber\\ &&T_{i,j}=2i \frac{\sqrt{p_0}(1-p_0)}{\cal
N}\sin(\phi_i-\phi_j)e^{2i\psi},\\
\label{def2} &&\,\nonumber\\
&&R^i_{j,k}=\frac{\sqrt{p_0}}{\cal N}
\Bigl[\bigl(e^{2i\phi_i}+p_0e^{2i(\psi-\phi_i-\phi_j-\phi_k)}\bigr)
\bigl(e^{2i\phi_j}+e^{2i\phi_k}\bigr)\nonumber\\
&&\hspace{2cm} +2+2p_0e^{2i\psi} \Bigr].\label{def3}
\end{eqnarray}
In Eqs. (\ref{def1})-(\ref{def3}), $\psi=\sum\phi_i$ is the net
phase; normalization factor is defined as
\begin{equation} \label{N} {\cal N}=
2+2p^2_0e^{2i\psi}
+p_0(e^{2i\phi_1}+e^{2i\phi_3})(e^{2i\phi_2}+e^{2i\phi_4}).
\end{equation}
One can check that the matrix Eq. (\ref{arbphase}) is unitary. For
arbitrary phases, $\phi_i$, it does not reduce to the form Eq.
(\ref{gsm}) with equal reflection, transmission, and deflection
probabilities for {\it all} directions of incident channels.
Global isotropy is restored upon averaging over $\phi_i$. For
example, if we choose a particular set, $\phi_1=\phi_2=\pi/4$,
$\phi_3=\phi_4=-\pi/4$, the matrix Eq. (\ref{arbphase}) assumes
the form
\begin{widetext}
\begin{equation}
\label{effective2}\hat{{\cal S}}_0= \frac1{1+p^2_0}
\left(\begin{array}{cccc} -\sqrt{p_0}(1+p_0)
&-\frac i{\sqrt{2}}(1-p_0)^2 &-i\sqrt{p_0}(1-p_0) &-\frac 1{\sqrt{2}}(1-p^2_0)\\
\,\\
\frac i{\sqrt{2}}(1-p_0)^2 &-\sqrt{p_0}(1+p_0)
 &\frac 1{\sqrt{2}}(1-p^2_0)&-i\sqrt{p_0}(1-p_0)\\
\,\\
-i\sqrt{p_0}(1-p_0)&-\frac 1{\sqrt{2}}(1-p^2_0)& -\sqrt{p_0}(1+p_0) &-\frac i{\sqrt{2}}(1-p_0)^2\\
\,\\
\frac 1{\sqrt{2}}(1-p^2_0)&-i\sqrt{p_0}(1-p_0) &\frac i{\sqrt{2}}(1-p_0)^2& -\sqrt{p_0}(1+p_0)\\
\end{array}\right).
\end{equation}
\end{widetext}
This choice favors deflection $|d_2|$ over deflection $|d_1|$. The
asymmetry between $|d_2|$ and $|d_1|$ is compensated by
realization in which $\phi_1=\phi_2=-\pi/4$,
$\phi_3=\phi_4=\pi/4$; for this realization, $|d_1|$ and $|d_2|$
switch places. It is seen from Eq. (\ref{arbphase}) that presence
of reflectors eliminates the singular character of junction matrix
${\cal S}_0$ by restoring finite forward and backward scattering
probabilities. In other words, the use of the matrix Eq.
(\ref{arbphase}) instead of node matrix Eq. (\ref{gsm}) would
reveal full localization of electron states at any $p_0$. However,
for small $p_0$, the parametric space of matrices Eq.
(\ref{arbphase}) is restricted to small transmission and
reflection, $|t|^2\sim |r|^2\sim p_0$, the domain where the mean
free path Eq. (\ref{mfp}) is $\sim 1$. Despite what  Eq.
(\ref{mfp}) predicts, we will get large values of localization
length, $\xi(p_0)$, in the domain $p_0\ll1$.

We complete our reformulation of the non-chiral network model by
observing that the random phases on the links between two
neighboring reflectors can be incorporated into $\phi_i$. This
allows one to combine the two reflectors on the same link into a
single effective $2\times 2$ scatterer on this link, as it is
shown in Fig. \ref{reform}b. The corresponding effective
scattering matrix,
\begin{equation} \label{Matrixp} {\cal P}=
\left(\begin{array}{cc}
\sqrt{1-p}&\sqrt{p}\\
\,\\
-\sqrt{p}&\sqrt{1-p}
\end{array}\right),
\end{equation}
has the same form as Eq. (\ref{Matrixp0}) with
\begin{equation} \label{twoone}
p= \frac{4p_0}{(1+p_0)^2}.
\end{equation}
The resulting network consisting of junctions at nodes and
effective scatterers on the links is shown in Fig. \ref{reform}b.
On the microscopic level, the node ${\cal S}_0$ corresponds to the
junction with confinement shown in Fig. \ref{junction}a, while the
link matrix ${\cal P}$ corresponds to point-contact confinement,
Fig. \ref{junction}b.

\section{Orbital action of a weak magnetic field}

As was discussed in the Introduction, weak  magnetic field in
which delocalization transition takes place, is already strong
enough to drive random phases on the links of the network Fig.
\ref{reform}b into the unitary class. We also need to incorporate
the Lorentz-force effect of magnetic field. For free electrons,
the Lorentz force  curves their trajectories. In the network Fig.
\ref{reform}b it affects the properties of junctions only, leading
to imbalance between deflection to the left and deflection to the
right. Note, that general  properties of a four-terminal junction
in magnetic field were previously studied in Refs.
\onlinecite{Ravenhall, Baranger, Beenakker} for various forms of
confinement potential in relation to experiments \cite{Timp87,
Roukes87, Chang88, Ford88, Simmons88} on the Hall quantization in
narrow channels.

In order to incorporate orbital action of magnetic field into the
general network Fig. \ref{network}, one has to place a weakly
chiral $S$-matrix into each node. A possible form of such
$S$-matrix is
\begin{eqnarray}
\label{chsm}
S_{ch}=\left(\begin{array}{cccc} r_1&d_2&t_3&D_4\\
D_1&r_2&d_3&t_4\\t_1&D_2&r_3&d_4\\d_1&t_2&D_3&r_4
\end{array}\right),
\end{eqnarray}
where the complex coefficients have absolute values
\begin{eqnarray} \label{par} &&|r_i|=r,\quad |D_i|=D,\quad
|d_i|=d,\quad |t_i|=t;\nonumber\\
&&r^2+D^2+d^2+t^2=1.
\end{eqnarray}
If magnetic field whirls electrons, say, to the right, one has
$D>d$. For example, one can choose the following realizations of
the above chiral matrix
\begin{eqnarray}
\label{wchm} S_{ch}=\!\left(\begin{array}{cccc}
r&-d-ib_-&t&d+ib_+\\
d+ib_+&r&-d+ib_-&t\\
t&-d+ib_+&-r&-d+ib_-\\
d+ib_-&t&d-ib_+&-r
\end{array}\right)\!,\nonumber\\
\end{eqnarray}
provided that $r$, $d$, $t$, and $b_{\pm}$ are positive real
numbers ($t>r$) with
\begin{eqnarray} \label{morepar}
\frac{b_-}{b_+}=\frac{t-r}{t+r},\quad r^2+t^2+2d^2+b^2_++b^2_-=1.
\end{eqnarray}
We see that the difference between the scattering probabilities to
the right and to the left is $|b_+|^2-|b_-|^2$. This difference is
non-zero, as follows from first identity in Eq. (\ref{morepar}).
Note that for the particular choice Eq. (\ref{chsm}), the ratio,
$r/t$, "controls" the magnetic field strength. Indeed, for $r=0$
we have $b_+=b_-$.

In general, dimensionless Hall resistance of a junction is
expressed via field-dependent elements of the matrix Eq.
(\ref{gsm}) as follows \cite{Ravenhall, Baranger, Beenakker}
\begin{equation}\label{HR}
R_H=\frac{2(d_1^2-d_2^2)}{(2t^2+d_1^2+d_2^2)^2+(d_1^2-d_2^2)^2}.
\end{equation}
Note that for the weakly chiral matrix Eq. (\ref{wchm}) $R_H$ is
the same for all directions of incidence. Naturally, the degree of
bending action of magnetic field is represented by the difference,
$(|d_1|^2-|d_2|^2)$. The advantage of our reformulation of the
network model, described in Sec. II, is that controlled chirality
can be incorporated into the network in a natural way  upon
replacement of the non-chiral matrix ${\cal S}_0$, Eq.~(\ref{S0}),
by
\begin{equation}
\label{Matrixq} {\cal S} =\left(
\begin{array}{cccc}
0&-{\scriptstyle\sqrt{1-q}}&0&{\scriptstyle -\sqrt{q}}\\
{\scriptstyle \sqrt{q}}&0&{\scriptstyle \sqrt{1-q}}&0\\
0&{\scriptstyle -\sqrt{q}}&0&{\scriptstyle \sqrt{1-q}}\\
{\scriptstyle \sqrt{1-q}}&0&{\scriptstyle -\sqrt{q}}&0
\end{array}\right).
\end{equation}
The matrix ${\cal S}$ is parameterized by a {\it single} number,
$q=d_1^2$, which varies between $q=0$ and $q=1$. The Hall
resistance is expressed via $q$ as follows:
\begin{equation}\label{HRq}
R_H=\frac{2q-1}{q^2+(1-q)^2}.
\end{equation}
It is an odd function of
\begin{equation}
\left(\frac12-q\right)\propto \omega_c,
\end{equation}
i.e., the difference $(1/2-q)$ can be viewed as a quantitative
measure of the magnetic field strength. Concluding this section,
the link matrix ${\cal P}$, Eq.~(\ref{Matrixp}), parameterized by
single parameter, $p$, and the node matrices ${\cal S}$,
Eq.~(\ref{Matrixq}), parameterized by single parameter, $q$, fully
define a minimal network model, Fig. \ref{reform}b. The advantage
of this network is that the disorder and the magnetic field can be
"tuned" independently by changing $p$ and $q$, respectively. We
will call this model a p-q model. Localization properties of this
model are studied below.

\section{p-q model: limit of strong disorder}

\begin{figure}[t]
\centerline{\includegraphics[width=90mm,angle=0,clip]{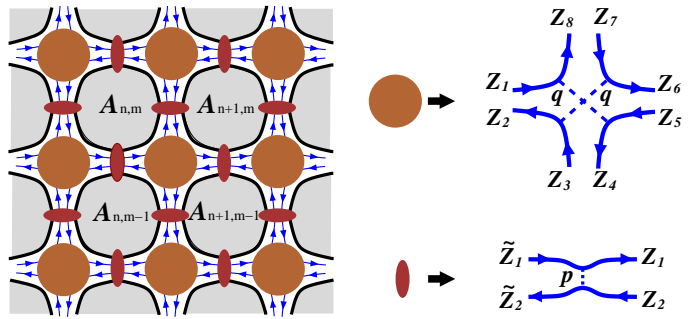}}
\caption{(Color online) Left: In two-channel p-q model, electron
motion is restricted to the spaces between forbidden regions
$A_{n,m}$. The centers of forbidden regions form a square lattice.
Point contacts on the links describe the backscattering by
disorder; bend-junctions at the nodes describe the orbital action
of magnetic field.  Right: Scattering matrices of the junction and
of the point contact. } \label{pqnetwork}
\end{figure}
In order to get a qualitative insight into the phase diagram of
the {\it quantum} p-q model, we start with artificial limit of
strong disorder. To define the strong disorder, note that in the
original p-q model the values of $p$ and $q$ are the same for all
junctions and point contacts. In other words, distribution
functions of the parameters $p$ and $q$ are
\begin{equation}\label{disf0}
f(p_i)=\delta(p_i-p),\quad f(q_j)=\delta(q_j-q).
\end{equation}
By a strong disorder we mean the following distribution of $p$ and
$q$,
\begin{eqnarray}\label{disf}
&&f(p_i)=p\,\delta(1-p_i)+(1-p)\delta(p_i),\\
&&f(q_j)=q\,\delta(1-q_j)+(1-q)\delta(q_j),\nonumber
\end{eqnarray}
so that scattering by the point contact and deflection at the
junction are still $p$ and $q$ {\it on average}. However, unlike
Eq. (\ref{disf0}), the point contact reflects {\it fully} in  $p$
percent of cases, and transmits fully in the rest $(1-p)$ percent
of cases. Similarly, according to Eq. (\ref{disf}), the junction
deflects only to the right in $q^2$ percent of cases, deflects
only to the left in $(1-q)^2$ percent of cases; in the remaining
$2q(1-q)$ percent of cases the deflection takes place both to the
left and to the right depending on incoming channel.

In considering a strong disorder, our motivation stems from the
original CC model, in which the nodes of the network are chiral
saddle points. If we introduce disorder in the transmission of
saddle points similar to Eq.~(\ref{disf}), then, at critical
energy, $50\%$ of saddle points will fully transmit, and $50\%$
will fully reflect. For such a disorder, quantum-mechanical
interference becomes irrelevant. However, the state will remain
critical \cite{Kivelson93}, separating the phases with
$\sigma_{xy}$ differing by $1$. In this limit of strong disorder,
quantum delocalization is replaced by classical percolation
transition occurring {\it at the same energy}. Our expectation,
which will be later supported by numerical simulations, is that
similar to CC model, considering the limit of strong disorder of
the p-q model will yield the positions of classical delocalization
transitions, which coincide with the positions of quantum
delocalized states in the original p-q model.
\begin{figure}[t]
\centerline{\includegraphics[width=60mm,angle=0,clip]{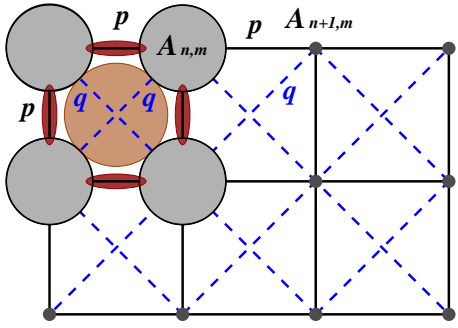}}
\caption{(Color online) Limit of strong disorder. The centers of
forbidden regions, $A_{n,m}$ and $A_{n,m-1}$, are connected by the
p- bond, while the centers of forbidden regions, $A_{n-1,m-1}$ and
$A_{n,m}$, are connected by a q- bond. Electron delocalization
reduces to the percolation on the lattice consisting of p- and q-
bonds.}
 \label{bondperc}
\end{figure}

The realization with $p_i=1$ corresponds to a "closed" point
contact, which reflects incoming waves from both directions.
Classically, presence of such a reflecting barrier can be
interpreted as a {\it bond} installed between the neighboring
forbidden regions, $A_{n,m}$ and $A_{n+1,m}$, of confining
potential, Fig. \ref{pqnetwork}. Below we will refer to this bond
as a p- bond. Similarly, we introduce q- bonds, installed between
the forbidden regions, $A_{n,m}$ and $A_{n\pm1,m\pm1}$ in Fig.
\ref{pqnetwork}. Then, deflection only to the right corresponds to
two crossed q- bonds, deflection only to the left corresponds to
the situation when all four forbidden regions $A_{n,m}$, defining
the junction, are disconnected. One right-diagonal q- bond
describes the situation when the junction deflects the fluxes
incident from the left and from the right channels to the left,
and fluxes incident from the up and down to the right. Similarly,
one left-diagonal q- bond signifies reflection from the left and
right channels to the right, while the up and down channels are
deflected by the junction to the left. Thus we arrive at the
percolation problem of connectivity of the forbidden regions
$A_{n,m}$ via p- and q- bonds, see Fig. \ref{bondperc}. Some
limits of this problem are transparent. For example, for small $p$
and $q$, the $A_{n,m}$ regions are mostly disconnected. Also, for
$p>1/2$, global connectivity exists for any $q$. It is intuitively
clear that adding small portion of q- bonds facilitates
connectivity and shifts the position of percolation transition
from $p=1/2$ to lower values of $p$. Quantitatively, we will
search for the boundaries of percolation transition on the $p,q$
plane by employing the real-space renormalization group approach
to the 2D percolation \cite{Stenley}.

\subsection{Real-space renormalization-group analysis of the percolation
problem}
\begin{figure}[b]
\centerline{\includegraphics[width=80mm,angle=0,clip]{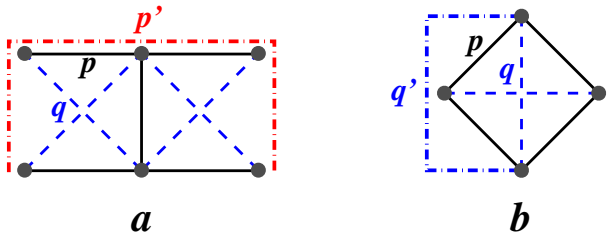}}
\caption{(Color online) a: Illustration of the real-space
renormalization-group procedure. The connectivity of a superbond
defined by five p- bonds is enhanced by q- bonds, Eq.
(\ref{largep}). b: Percolation picture at small $p$. p-bonds
"assist" the connectivity of q- bonds, Eq. (\ref{smallp}).}
 \label{supbond}
\end{figure}

\begin{figure}[t]
\vspace{-0.3cm}
\centerline{\includegraphics[width=90mm,angle=0,clip]{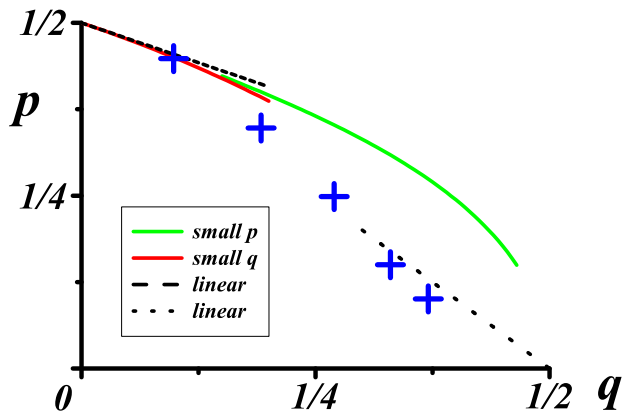}}
\caption{(Color online) The line of percolation transitions in the
$(p,q)$-plane is plotted from Eq. (\ref{largep}) for small $q$
(red line) and from Eq. (\ref{smallp}) for small $p$ (green line).
Dashed and dotted straight lines are the asymptotes, Eq.
(\ref{smallqbound}) and  Eq. (\ref{linear}) with $\alpha=1$,
respectively. Crosses show the positions of quantum delocalization
transition inferred from quantum simulations for five values of
``energies'', $p$. } \label{phasediag}
\end{figure}

The original approach of Ref. \onlinecite{Stenley} applies when
$q=0$, i.e., when only p- bonds connecting the sites of $A_{n,m}$
square lattice are present. Within this approach, five p- bonds
are replaced by one superbond, as shown with full lines in Fig.
\ref{supbond}a. Probability, $p^\prime$, that superbond is
present, is expressed via probability that p- bond is present, as
\begin{eqnarray}\label{supconnect}
p^\prime\!\!&=&\!\! P_5(p)+ P_4(p) + P_3(p) + P_2(p)+P_1(p)+P_0(p),\nonumber\\
&=&\!\!f_0(p),
\end{eqnarray}
where $P_n(p)$ is the partial probability that the superbond is
present when $n$ original bonds are present. A simple counting of
variants yields
\begin{eqnarray}\label{Pn}
&&P_5(p)=p^5,\quad P_4=5p^4(1-p),\quad P_3= 8p^3(1-p)^2,\nonumber\\
&&P_2(p)=2p^2(1-p)^3,\quad P_1=0,\quad P_0=0.
\end{eqnarray}
Remarkable feature of the transformation Eq. (\ref{supconnect}) is
that its fixed point, $p^\prime=p$, coincides with the exact bond
percolation threshold, $p=1/2$, on the square lattice. The q-
bonds, shown with dashed lines in Fig. \ref{supbond}. We will
incorporate them into renormalization group transformation
assuming that their role is the enhancement of connectivity of the
superbond. This enhancement occurs differently depending on how
many original bonds are present. For example, if all five or four
bonds are present, the connectivity of superbond is guaranteed
even without any q- bonds, so that $P_5$ and $P_4$ are not
affected by q- bonds. When three p- bonds are present, there are
two variants when superbond does not connect. Then the probability
that it connects {\it in the presence} of q- bonds is given by
\begin{equation}
\label{P3pq} P_3(p,q)=P_3(p)+2p^3(1-p)^2\bigl[2q(1-q)+q^2\bigr].
\end{equation}
The second term is the product of probabilities that p-
connectivity is absent and that q- bonds restore it. The factor in
the square brackets accounts the fact that restoration can happen
by installing one q- bond (two variants) as well as two q- bonds
(one variant). The expression for $P_2(p,q)$ has a similar
structure,
\begin{equation}
\label{P2pq}
P_2(p,q)=P_2(p)+p^2(1-p)^3\bigl[2\cdot4\bigr(2q(1-q)+q^2\bigl)\bigr].
\end{equation}
Overall, there are ten configurations when two p- bonds of the
superbond are present. Out of these ten, there are two variants
when the superbond connects; in the remaining eight variants the
superbond does not connect. The factor in the square brackets in
Eq. (\ref{P2pq}) describes the probability that in these eight
variants q- bonds make the superbond connect. Note that the $q$
dependence of the second term in Eq. (\ref{P3pq}) is the same as
in the case of $P_2(p,q)$. This reflects the fact that in both
cases installing either one or two q- bonds restore connectivity.
If there is only one p- bond, it can be either "vertical" (one
variant) or horizontal (four variants). In the first case,
probability of restoring connectivity is  a product of the
probabilities that it is restored both "to the left" and "to the
right" from the p- bond. If the p- bond is horizontal, there are
three q- bonds that might participate in the restoration of the
connectivity. This yields
\begin{eqnarray}
\label{P1pq} &&P_1(p,q)=p(1-p)^4\bigl[
(2q(1-q)+q^2\bigl)^2\nonumber\\
&&+4\bigl(q(1-q)^2+3q^2(1-q)+q^3\bigr)\bigr].
\end{eqnarray}
Finally, $P_0(p,q)$ is the probability that the superbond connects
via q- bonds only. All four q- bonds can participate in
restoration, see Fig. \ref{supbond}a. In particular, the
connectivity can go through the upper middle site, lower middle
site, or both, resulting in
\begin{equation}
\label{P0pq} P_0(p,q)=(1-p)^5\bigl[
2q^2(1-q)^2+4q^3(1-q)+q^4\bigr]
\end{equation}
The net probability that the superbond connects is the sum of
$P_5(p)+ P_4(p)$ and Eqs. (\ref{P3pq})- (\ref{P0pq}), namely
\begin{eqnarray}
\label{largep} &&f(p,q)=f_0(p)+2 p^3 (1 - p)^2 \bigl[2 q -
q^2\bigr]\\
&&+8p^2(1-p)^3\bigl[2 q -
q^2\bigr]\nonumber\\
&&+p(1-p)^4\bigl[4q+8q^2-8q^3+q^4\bigr]+
(1-p)^5\bigl[2q^2-q^4\bigr].\nonumber
\end{eqnarray}
We determine the line of percolation transitions on the $p,q$-
plane upon equating $f(p,q)$ to $\frac{1}{2}$. Solution of this
equation is plotted in Fig. \ref{phasediag}. At small $q$ the
$p(q)$ boundary is linear,
\begin{equation}\label{smallqbound}
p_c(q)=\frac12-\frac6{13}\,q.
\end{equation}

In the above procedure we assumed that the effect of q- bonds is a
correction to the percolation over p- bonds. Consider now the
opposite limit, where $q$ is close to $1/2$ and $p$ is small, so
that the percolation is dominated by the q- bonds, while the p-
bonds constitute a small correction. First we note that for $p=0$,
the regions $A_{n,m}$ with $n+m$ even and $n+m$ odd are {\it
decoupled}. Moreover, $q=1/2$ corresponds to bond percolation
threshold in both decoupled sublattices. Now adding small portion
of p- bonds facilitates percolation for $q<1/2$. To describe this
facilitation quantitatively we turn to Fig. \ref{supbond}b. This
figure illustrates that, instead of one missing q- bond, a {\em
pair of one horizontal and one vertical} p- bonds can provide the
connection, and that there are two such variants. Resulting shift
of the threshold position is determined by the condition
\begin{equation}
\label{smallp} f(p,q)=q^{\prime}=q+2(1-q^2)p^2 =\frac12,
\end{equation}
where $q^{\prime}$ is the renormalized probability that q- bond
connects. The origin of the factor $(1-q^2)$ in Eq. (\ref{smallp})
is the following. If the vertical q- bond is missing [with
probability $(1-q)$], the two p- bonds which restore connectivity
can either share a common site (with probability $2p^2$), or two
p- bonds can be connected to each other via a horizontal q- bond
(with the probability $2p^2q$). The sum of probabilities of these
two realizations should be multiplied by $(1-q)$.  The resulting
from Eq. (\ref{smallp}) $p(q)$ dependence,
\begin{equation}\label{smallpbound}
p_c(q)=\sqrt{\frac{1-2q}3}\,,
\end{equation}
is plotted in Fig. \ref{phasediag}. In fact, as seen from Fig.
\ref{phasediag}, the asymptotic behaviors found from Eqs.
(\ref{largep}) and (\ref{smallp}) match very closely near
$q=\frac{1}{4}$.

Eq. (\ref{smallp}) describes the situation when percolation occurs
in one, e.g., $n+m$ even sublattice, while the $n+m$ odd
sublattice paly an auxiliary role. This picture is violated near
the degeneracy point $p=0$, $q=1/2$. Upon approaching to this
point, {\it both} sublattices should be treated on the equal
footing. Moreover, the position of the boundary on the $p,q$ plane
is governed not by a local arrangement of the bonds but rather by
large-scale behavior of the clusters in both sublattices. We will
discuss percolation transitions in this region in a separate
subsection below, after we establish the relation between
percolation and electron trajectories in magnetic field.

\subsection{Implication of the percolation transition for
transport}

The prime question is: to what extent the connectivity of
$A_{i,j}$ regions studied in the previous subsection governs the
transport over the regions located {\it between} the forbidden
regions $A_{i,j}$. To emphasize that this question is non-trivial,
note that, while p- and q- bonds facilitate the connectivity, a
combination of two p-bonds at a given junction and one q- bond
through the same junction can give rise to {\it localized}
electron trajectory, circling around the junction.
\begin{figure}[t]
\centerline{\includegraphics[width=90mm,angle=0,clip]{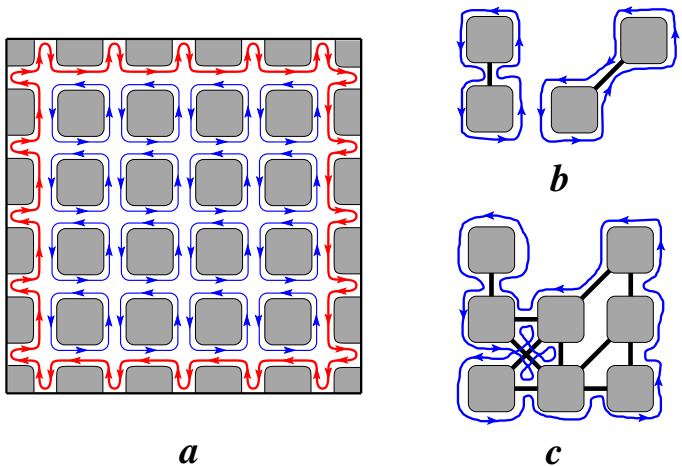}}
\caption{(Color online) a: Structure of electron trajectories at
$p=q=0$. Trajectories either encircle forbidden regions,
$A_{n,m}$, (brown lines), or constitute a chiral edge state due to
reflection from the boundaries (red line). b: Installing one p-
(vertical) or q- (diagonal) bond creates a trajectory encircling
two forbidden regions counter-clockwise. c: Example of a finite
cluster of p- and q- bonds. Formation of trajectory encircling the
hull.} \label{trajectories}
\end{figure}

To establish the connection between transport and percolation, we
start with the simplest case $p=0$, $q=0$.
It is apparent from Fig. \ref{pqnetwork} (see also Fig.
\ref{trajectories}a) that in this case electron trajectories are
closed counter-clockwise loops around forbidden regions $A_{i,j}$.
In other words, electron executes counter-clockwise motion along
the {\it perimeters} of $A_{i,j}$. Now let us switch on a single
p- bond, say, between $A_{n,m}$ and $A_{n,m-1}$, Fig.
\ref{trajectories}b. It is easy to see that installing this bond
creates a closed trajectory encircling two regions, joined by the
bond, in such a way
that the forbidden joined region
remains on the left. Thus, from two trajectories along the
perimeters of disconnected regions we get one trajectory {\it
along the perimeter of the joined region} Fig.
\ref{trajectories}b.
The same happens upon installing a single q- bond, as shown in
Fig. \ref{trajectories}b. In fact, this evolution is general:
if a portion of p- and q- bonds connect several forbidden regions
into a cluster, there appears a closed trajectory along outer
perimeter of the cluster. While moving along this {\it perimeter
trajectory}, the cluster remains on the left.
For the transport properties of p-q model, the outer perimeter,
which signifies the most delocalized trajectory existing with the
given cluster, plays a central role. Note that in the percolation
theory such a perimeter
is called a {\it hull}. Thus we see that hulls of the bond
percolation Fig. \ref{trajectories}c correspond to the most
important electron trajectories of the p-q model, in the sense,
that, upon approaching the percolation threshold, the hulls of big
clusters  join into even bigger hulls. Extent of the region
available for electron motion is determined by the size of the
typical cluster, which is the localization radius of the classical
percolation. While locally the motion occurs along the boundaries
of forbidden regions $A_{i,j}$, at large scales the hulls define
the extent of the motion. Now we can identify the point of the
percolation threshold at which $A_{i,j}$ get connected into
infinite cluster with the point when the hull trajectories become
infinite and connect opposite sides of macroscopic sample.
\begin{figure}[t]
\centerline{\includegraphics[width=90mm,angle=0,clip]{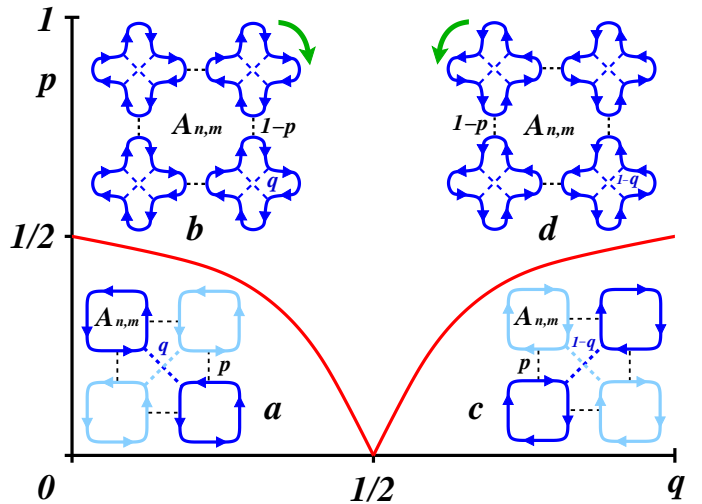}}
\caption{(Color online) Phase diagram (red line) of the p-q model
in the regime of strong disorder. Phases (a) and (c): electron
predominantly encircles forbidden regions $A_{n,m}$, moving
counter-clockwise (a) or clockwise (c). Coupling of neighboring
forbidden regions via p- and q- bonds is weak. Phases (b) and (d):
electron predominantly encircles nodes, moving counter-clockwise
(d) or clockwise (b). Coupling of nodes due to absent p- bonds is
weak. Edge trajectory disappears upon crossing the boundary (red
line), (a)~$\rightarrow$~(b) or (c)~$\rightarrow$~(d).}
 \label{diffph}
\end{figure}


Essentially, hulls can provide a connectivity through an infinite
sample exactly at the percolation threshold. Below and above this
threshold, hulls either do not provide a macroscopic connectivity
or establish a trajectory extending along the macroscopic edges of
a sample, depending on boundary conditions. Being translated into
the p-q model transport properties, this means that a non-zero
diagonal conductivity, $\sigma_{xx}\neq 0$, is possible only at
the percolation transition, which separates two insulating regimes
with $\sigma_{xx}= 0$.

We now choose the boundaries of the macroscopic p-q sample passing
through the centers of the end regions $A_{i,j}$, as shown in Fig.
\ref{trajectories}a, and assume full reflection at the boundaries.
It can be seen that for $p=0$, $q=0$, there is a macroscopic
trajectory spanning near the edges around the sample in the
clockwise direction. Therefore, in the phase $a$,
Fig.~\ref{diffph}, we have nonzero $\sigma_{xy}$, while
$\sigma_{xx}= 0$ because there is no trajectory through the sample
at $p=q=0$ and in the vicinity.

Previous consideration pertains to small enough $p$ and $q$. Let
us now move along the boundary $q=0$. In the absence of q- bonds
this corresponds to increasing connectivity, $p$, in conventional
bond percolation problem. It is easy to see that the edge state
disappears when we pass the percolation threshold $p=1/2$. Above
this point, the electron trajectories are strongly localized
around the junctions (phase $b$ in Fig. \ref{diffph}), and
describe the {\it clockwise} motion, unlike counter-clockwise
loops in the phase $a$.

Now we fix $p$ in the domain $(1-p)\ll 1$ and move along the q-
axis. At this point we note that the p-q model possesses a {\it
duality} $q\rightarrow (1-q)$. One transparent way to reveal this
duality is to trace the change in strongly localized trajectories,
phase $b$, upon changing $q$ by $(1-q)$. One can see that
trajectories remain unchanged, while the direction changes from
counterclockwise to clockwise. Such a change could be expected on
purely physical grounds, since transformation $q\rightarrow (1-q)$
means that the difference, $(1/2-q)$, which represents magnetic
field, changes sign. Naturally, reversal of magnetic field results
in the change of direction in which the $A_{n,m}$ regions are
circumvented. Therefore there is one-to-one correspondence between
the strongly localized phase $b$ and phase $d$. The same argument,
change of direction of the rotation as a result of $q\rightarrow
(1-q)$ transformation, suggests that the phase $c$ is mirror image
of the phase $a$, with the opposite sign of $\sigma_{xy}$.

\subsection{ Vicinity of the degeneracy point  $p=0$, $q=1/2$}

\begin{figure}[t]
\centerline{\includegraphics[width=80mm,angle=0,clip]{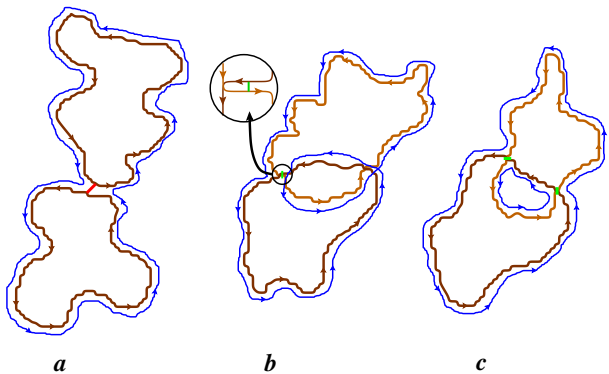}}
\caption{(Color online) Vicinity of the point $p=0$, $q=1/2$, of
the phase diagram Fig. \ref{diffph}. a: Two q-clusters on the {\it
same} sublattice hybridize upon installing a q- bond.  b: A q-
cluster of $n+m$ odd sublattice (upper) and a q- cluster of $n+m$
even sublattice (lower) overlap. A joined trajectory (thin blue
line) is formed upon installing of a {\it single} p- bond. The
blowup illustrates hybridization of trajectories on a microscopic
level. c: Same as (b), but with two p- bonds connecting q-
clusters. Blue lines illustrate that the hull trajectory and
"internal" trajectory are disconnected.} \label{clusters}
\end{figure}

In the close vicinity, $(1/2-q)\ll 1$, $p\ll 1$, of the degeneracy
point, one has big clusters of q- bonds in $(n+m)$ even and
$(n+m)$ odd sublattices, which are statistically equivalent. Let
us start our consideration from some $q=q_0<1/2$ and $p=0$.
Connectivity via q- bonds can be enhanced either by adding p-
bonds or by shifting $q$ closer to $1/2$. We argue that, close
enough to the degeneracy point, both operations are equivalent.
This is because the number of "even" q- bonds, i.e., the q-bonds
from the $(n+m)$-even sublattice, and the number of "odd" q-
bonds, i.e., the q-bonds from the $(n+m)$-odd sublattice, involved
in a typical cluster, is the same. Once this "equal participation"
of two sublattices is achieved, it is preserved upon increasing
both $p$ and $q$. The value of $p$ necessary to achieve this
equal-participation regime obviously depends on proximity of $q$
to $1/2$. A way to estimate this necessary $p$ is based on the
following reasoning. The spatial separation between neighboring p-
bonds is $\sim 1/\sqrt{p}$. If this separation is smaller than the
typical size of q- cluster, $\sim|1/2-q|^{-4/3}$, on a given
sublattice, i.e.,
\begin{equation}\label{est}
p>\left(\frac12-q\right)^{8/3},
\end{equation}
then p- bonds connect clusters from different subnetworks. The way
in which equal-participation regime sets in is illustrated in Fig.
\ref{clusters}. In Fig.~\ref{clusters}a the condition
Eq.~(\ref{est}) is not met; clusters grow independently in each
sublattice as $q$ increases. Figs. \ref{clusters}b,c illustrate
that a typical clusters on one sublattices is overlapped by some
other typical cluster from another sublattice. Therefore, few p-
bonds per cluster, see Eq.~(\ref{est}), are sufficient for
formation of a unified cluster. Percolation boundary, $p_c(q)$,
lies above the boundary of equal-participation regime
Eq.~(\ref{est}). It is reasonable to assume that the percolation
threshold corresponds to certain portion of p- {\it and} q- bonds
per site. Having in mind that in equal-participation regime p- and
q- bonds are equivalent within a factor, the above assumption
leads us to
\begin{equation}\label{linear}
p_c(q)\simeq \alpha{\Big |}\frac12-q{\Big |},
\end{equation}
with numerical coefficient, $\alpha\sim 1$. This linear behavior,
as well as the forms Eqs. (\ref{smallqbound}) and
(\ref{smallpbound}), are in agreement with the condition Eq.
(\ref{est}), which in fact should hold in the whole domain where
we expect delocalization transitions with the participation of p-
bonds. The above reasoning, leading to the linear boundary Eq.
(\ref{linear}) is by no means rigorous. We were able to come up
with more compelling reasoning that for the case when half of q-
bonds are replaced by $1-q$, the percolation transition at small
$p$ indeed occurs at $p\sim(1/2-q)$. Note however, that one cannot
claim that this auxiliary problem provides evidence for linear
boundary in our p-q model. This is because approaching the point
$(p,q)=(0,1/2)$, can depend of the direction of the approach, so
that the approach along a path with "balanced" $q$ and $1-q$ bond
can yield a different boundary.

Moreover, this linear dependence is a source of an apparent
doubling of the critical exponent when one approaches the
delocalization point along the horizontal line
$p=\text{const}\ll1$.

\subsection{Doubling of the critical exponent}

The shape of the percolation transition line Eq. (\ref{linear})
for which the condition Eq. (\ref{est}) is met, allows one to make
quantitative predictions about the behavior of the localization
radius, $\xi(p,q)$. The reasoning goes as follows. Eq. (\ref{est})
ensures that the typical clusters on the two sublattices with
sizes $\xi(q)$ are connected by p- bonds. Now, if we keep $q$
unchanged and approach the critical line, $p_c(q)$, from below,
the localization radius grows as
\begin{equation}\label{doubling}
\xi(p,q)\sim\frac{\xi(q)}{\bigl[p_c(q)-p\bigr]^{4/3}}.
\end{equation}
While this relation is asymptotically exact near the
delocalization line, we assume that it still holds deeper into the
region (a), Fig. \ref{diffph}. Then, in order to find an explicit
dependence $\xi(p,q)$, Eq. (\ref{linear}) can be used in Eq.
(\ref{doubling}), yielding
\begin{equation}\label{expdep}
\xi(p,q)\sim\frac{\xi(q)}{\left[\alpha\left(\frac12-q\right)-p\right]^{4/3}}.
\end{equation}
Note now that Eq. (\ref{expdep}) can be interpreted as a {\it
doubling} of the critical exponent. As one starts at some point,
$(p_0,q)$, such that $p_0\ll1$ and $(1/2-q)\sim 1$ and move
towards the transition boundary along the horizontal line $p=p_0$,
localization radius changes as follows
\begin{equation}\label{expdoubling}
\xi(p_0,q){\Big
|}_{q<1/2}\sim\frac{1}{\left[\frac12-q\right]^{4/3}\left[\frac12-\frac
{p_0}\alpha-q\right]^{4/3}}.
\end{equation}
For $(1/2-q)\gg p_0/\alpha$ Eq. (\ref{expdoubling}) assumes the
form
\begin{equation}\label{doubled}
\xi(p_0,q)\sim\frac1{\left[\frac12-q\right]^{8/3}},
\end{equation}
which means that the subcritical behavior of $\xi(p_0,q)$ as a
function of $q$ is similar to the usual $|q-q_c|^{-4/3}$, but with
the doubled critical exponent. Therefore, it is the linear
behavior Eq. (\ref{linear}) which leads to the doubling. On the
other hand, as we have reasoned in the previous subsection, the
linearity of the boundary of the percolation transition is a
consequence of the "interaction" of two subnetworks.

The remaining question is the behavior of $\xi(p,q)$ along the
line $q=1/2$. To address this question, note that there are {\it
two} delocalization transitions which take place as $q$ is changed
along the line $p=p_0\ll1$; the first one at $q=q_{c1}<1/2$ and
the second at $q=q_{c2}>1/2$. By virtue of duality, $q_{c1}$ and
$q_{c2}$ are related as $q_{c1}+q_{c2}=1$. For small $p$, from Eq.
(\ref{linear}) we have
\begin{equation}\label{qc1}
q_{c1}(p)=\frac12-\frac p\alpha,\qquad q_{c2}(p)=\frac12+\frac
p\alpha.
\end{equation}
In the domain $q>1/2$ the $\xi(p,q)$ dependence can be found using
the above arguments
\begin{equation}\label{expdoubling1}
\xi(p_0,q){\Big
|}_{q>1/2}\sim\frac{1}{\left[q-\frac12\right]^{4/3}\left[q-\frac12-\frac
{p_0}\alpha\right]^{4/3}}.
\end{equation}
Recall that in Eq. (\ref{expdoubling}), the factor
$[1/2-q]^{-4/3}$ appears as a length of the unit plaquette in the
percolation over p-bonds. Then the second factor, $[1/2-
p_0/\alpha -q]^{-4/3}$, can be regarded as  a localization radius
associated with the delocalization transition at $q_{c1}(p_0)$.
The general expression which respects the duality and contains as
limits Eqs. (\ref{expdoubling}) and (\ref{expdoubling1}) reads
\begin{equation}\label{prodxi}
\xi(p,q)\sim \frac{1}{{\big |}\frac12-\frac{p}\alpha-q{\big
|}^{4/3} {\big |}\frac12+\frac{p}\alpha-q{\big |}^{4/3}}.
\end{equation}
In particular, along the  line $q=1/2$ this expression predicts
the following behavior of localization radius
\begin{equation}\label{center}
\xi(p){\Big |}_{q=1/2}\propto\frac1{p^{\,8/3}}.
\end{equation}

\subsection{Consequences of the doubling}

Below we will demonstrate numerically that Eq. (\ref{prodxi})
applies also to the quantum delocalization upon replacement $4/3$
by the quantum critical exponent of localization radius. Here we
would like to emphasize the similarity between the quantum version
of Eq. (\ref{prodxi}) and energy dependence of the localization
radius in the case of close, {\it e.g.}, spin-split delocalized
states in strong magnetic field. Behavior of $\xi$, similar to Eq.
(\ref{prodxi}), was conjectured in Ref. \onlinecite{Doubling} and
demonstrated numerically in Refs.
\onlinecite{LeeChalker94,LeeChalkerKo94}. In our case, two close
delocalized states correspond to {\it opposite} directions of
magnetic field, and thus have opposite chiralities. On the
contrary, in Refs. \onlinecite{Doubling,
LeeChalker94,LeeChalkerKo94,hanna95}, chiralities of both
delocalized states, which are close in energy, are {\it the same}.
More microscopic demonstration of doubling in the system with two
close in energy delocalized states with the same chirality can be
found in Ref. \onlinecite{Gramada97}. The situation considered in
Ref. \onlinecite{Gramada97} was two layers with smooth random
potential coupled by tunneling with amplitude, $t_0$. For a single
layer, the transmission of a saddle point is given by
$T(E/\Gamma)=[1+\exp(E/\Gamma)]^{-1}$. For two layers, tunneling
allows to bypass saddle points. Instead, the role of a saddle
point is played by the region where equipotentials from different
layers come close to each other, but do not intersect. It was
demonstrated in Ref. \onlinecite{Gramada97} that for $E$ outside
the interval $(-t_0,t_0)$ the transmission of such an effective
saddle point is given by
$\tilde{T}(E/\tilde{\Gamma})=T(E^2/\tilde{\Gamma}^2)$, where
$\tilde{\Gamma}$ depends on $t_0$ as $t_0^{1/4}$. Therefor, while
in the vicinity of delocalized states $E=\pm t_0$ the critical
behavior of localization length is $\xi\propto |E\pm t_0|^{-\nu}$,
outside the interval $(-t_0,t_0)$ we have $\xi\propto
(E^2)^{-\nu}$, which mimics the doubling of the critical exponent.

\subsection{Physical interpretation of the phase diagram Fig.
\ref{diffph}}

In physical terms, the p-q boundary in Fig.~\ref{diffph} relates
the backscattering probability, $p$, and magnetic filed,
$(1/2-q)$, at which delocalization transition takes place. The
parameter $p$ can be viewed as a measure of disorder and also as a
measure of {\it electron energy}, $E_{\scriptscriptstyle F}$;
obviously, $p$ decreases monotonously with increasing
$E_{\scriptscriptstyle F}$. Then, the domain $q<1/2$ of the phase
boundary in Fig.~\ref{diffph} translates into the low-field
region, $\omega_c\tau<1$, of the dependence $E_0(\omega_c)$ in
Fig.~\ref{levit}. Correspondingly, the domain $q>1/2$ maps onto
$E_0(\omega_c)$ dependence with reversed sign of the magnetic
field. More detailed correspondence between Figs.~\ref{levit} and
\ref{diffph} can be established upon identification of
$\sigma_{xy}$ values for different phases. As we demonstrated
above using percolation language, there is no edge state in the
central phase, which includes regions $b$ and $d$ in Fig.~\ref{diffph}.
Since this phase represents the region
$E_{\scriptscriptstyle F}< E_0(\omega_c)$ in Fig. \ref{levit},
this phase should be identified with the Anderson insulator. We
have also demonstrated, see Fig.~\ref{trajectories}a, that there
is one chiral edge state in the phase $a$ of the phase diagram.
Thus, in the region $E_{\scriptscriptstyle F}> E_0(\omega_c)$ in
Fig.~\ref{levit} we have $\sigma_{xy}=1$, so this phase is a
quantum Hall insulator. In the region $p>1/2$ there are no
delocalized states in Fig. \ref{diffph}. This region translates
into the strongly localized regime, $E_{\scriptscriptstyle F}<
1/\tau$, i.e., below the minimum of the $E_0(\omega_c)$ curve.

Linearity of the $p_c(q)$ boundary can be rewritten in terms of
observables. Upon identifying $p$ with $1/(k_Fl)$ and $(1/2-q)$
with $\omega_c\tau$, the position of the boundary can be presented
as $(k_Fl)(\omega_c\tau)=\text{const}$. This quantifies the
levitation rate, $\omega_c\tau\propto1/(k_Fl)$. Curiously, it
coincides with the prediction of scaling theory, Eq.
(\ref{positions}).

For conclusive confirmation of the levitation scenario it should
be demonstrated that the p-q phase boundary retains its shape in
the presence of quantum interference. In fact, the quantum and
"percolation" boundaries almost coincide. The evidence for that
will be presented in the next Section.

\begin{figure}[t]
\centerline{\includegraphics[width=50mm,angle=0,clip]{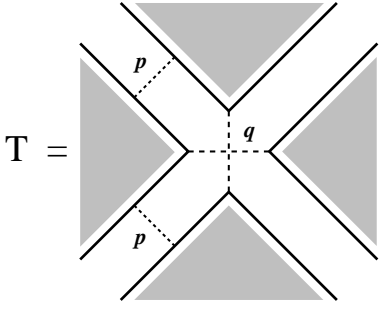}}
\caption{Illustration of the transfer matrix, $\text{\large T}$,
Eq.~(\ref{Tmatrix}), of the p-q model. Two q- bonds correspond to
the first matrix in the product Eq.~(\ref{Tmatrix}); two p- bonds
correspond to the third block-diagonal matrix in the product. }
\label{Tmatpic}
\end{figure}

\section{Numerical results for quantum delocalization}

\subsection{Transfer matrix}

Scattering matrices on the links and at the nodes are given by
Eqs. (\ref{Matrixp}) and (\ref{Matrixq}), respectively. With
regard to numerical simulations, the p-q model is quite similar to
the models with mixing of two co-propagating channels on the
links, studied in Refs. \onlinecite{Kagalovsky95,
kagalovsky97,LeeChalker94,LeeChalkerKo94}.

The transfer matrix, $\text{\large T}$, at each node of the
network is a $4\times 4$ matrix, which transforms four amplitudes
on the left into four amplitudes on the right. We incorporate the
link p-matrices and junction q-matrices into $\text{\large T}$ in
the way illustrated in Fig. \ref{Tmatpic}. It follows from   Fig.
\ref{Tmatpic} that $\text{\large T}$ matrix can be parameterized
as
\begin{widetext}
\begin{equation}
\label{Tmatrix} \text{\large T}=\! \left(\!\!
\begin{array}{cccc}
{\scriptstyle\frac1{\sqrt{1-q}}}&0&0&{\scriptstyle \sqrt{\frac q{1-q}}}\\
0&{\scriptstyle \frac1{\sqrt{q}}}&{\scriptstyle \sqrt{\frac{1-q}q}}&0\\
0&{\scriptstyle \sqrt{\frac{1-q}q}}&{\scriptstyle \frac1{\sqrt{q}}}&0\\
{\scriptstyle \sqrt{\frac q{1-q}}}&0&0
&{\scriptstyle\frac1{\sqrt{1-q}}}
\end{array}\right)\!\!\!
\left(\!\!
\begin{array}{cccc}
e^{i\varphi_1}&0&0&0\\
0& e^{i\varphi_2}&0&0\\
0&0&e^{i\varphi_3}&0\\
0&0&0& e^{i\varphi_4}
\end{array}\!\!\right)\!\!
\left(\!\!
\begin{array}{cccc}
{\scriptstyle\frac1{\sqrt{1-p}}}&{\scriptstyle \sqrt{\frac p{1-p}}}&0&0\\
{\scriptstyle \sqrt{\frac p{1-p}}}
&{\scriptstyle\frac1{\sqrt{1-p}}}&0&0\\ 0&0&
{\scriptstyle\frac1{\sqrt{1-p}}} &{\scriptstyle \sqrt{\frac p{1-p}}}\\
0&0&{\scriptstyle \sqrt{\frac p{1-p}}}
&{\scriptstyle\frac1{\sqrt{1-p}}}
\end{array}\!\!\right)\!\!\!
\left(\!\!
\begin{array}{cccc}
e^{i\phi_1}&0&0&0\\
0& e^{i\phi_2}&0&0\\
0&0&e^{i\phi_3}&0\\
0&0&0& e^{i\phi_4}
\end{array}\!\!\right).
\end{equation}
\end{widetext}
The second and the fourth matrices in the product describe random
phases acquired between the point contact and the junction.

Once the transfer matrix is specified, one can find the transfer
matrix of a slice with $M$ nodes in transverse direction. As
usual, to study the critical properties, numerical simulations are
performed on a system with fixed width $M=2^k$, then repeated for
$4$,$5$, $6$, $7$ and sometimes even $k=8$. By multiplying
transfer matrices for a stripe with $N$ slices and diagonalizing
the resulting total transfer matrix, it is possible to extract the
smallest positive Lyapunov exponent $\lambda_{M/2}$ (the
eigenvalues of the transfer matrix are $\exp(\lambda_i N)$). The
localization length, $\xi_M$, is proportional to
$1/\lambda_{M/2}$. The state is identified as a critical one when
the renormalized localization length, $\xi_M/M$, becomes
independent of the system width, $M$. We consider this as a
criterion for a transition point, that determines the value
$q=q_c$, for a fixed value of $p$. Practically, $q=q_c$ emerges as
a maximum in $\xi_M/M$ vs. $q$ for large system widths, when
finite-size errors become small.

Close to $q=q_c$, the ratios $\xi_M/M$ should satisfy a
one-parameter scaling
\begin{equation}
\frac{\xi_M}{M} =f\left(\frac{\xi (q)}M\right) , \label{nmscaling}
\end{equation}
which is commonly used to infer the localization length,~$\xi$.

We first check that our numerical data support
Eq.~(\ref{nmscaling}); to do so we fit all the data points onto
one curve according to Eq. (\ref{nmscaling}). The fit is carried
out with the help of a special optimization program. This
optimization program runs different critical $q$-values and
critical exponents $\nu$ and chooses the optimal sets, $q_c$ and
$\nu$, which provides the best agreement with Eq.
(\ref{nmscaling}).

We now briefly describe the optimization procedure. The routine
determines least-squares polynomial approximation by minimizing
the sum of squares of  the deviations of the data points from the
corresponding values of a polynomial. The argument of the function
fitted by the Chebyshev polynomials is $M|q-q_c|^\nu$. To choose
the optimal pair, $q_c$ and $\nu$, we run this routine for the
wide range of values  $q_c$ and $\nu$. For most of the points on
the critical line the values of $q_c$, obtained by two methods:
({\it i}) searching for $q$ at which $\xi_M/M$ is constant and
({\it ii}) using optimization procedure, agree with each other.

Note however, that in two limiting cases, $p\rightarrow 0$ and
$p\rightarrow 1/2$, where the data strongly fluctuate, apparent
discrepancies arise. For very small values of $p$ the off-diagonal
terms in the transfer matrix are close to $0$ (they are
$\sim\sqrt{p}$), leading to strong fluctuations in numerical
results. Physically, enhancement of fluctuations near $p=0$ is a
result of proximity to {\it two} critical points, $q_c$ and
$1-q_c$, see Fig. \ref{diffph}, where the doubling of critical
exponent takes place. On the other hand, when $p$ is close to
$1/2$, we have $q_c$ close to $0$. Strong fluctuations in the data
in this case is a consequence of small denominators, $1/\sqrt{q}$,
in the transfer matrix Eq. (\ref{Tmatrix}). Altogether, both
methods yield close values of $q_c(p)$.

The last remark on simulation procedure is on the boundary
conditions in transverse direction. In CC model, the periodic
boundary conditions in transverse direction are insured upon
imposing requirement on the structure of transfer matrix of even
slices only. The same is true for the p-q model. Unlike the CC
model, where the reflection and transmission at the nodes
alternate between subsequent slices, the elementary transfer
matrices in the p-q model, Eq. (\ref{Tmatrix}), are the same for
even and odd slices. This equivalence is a result of symmetry of a
single node with respect to $90^\circ$ rotations in the p-q model.

\subsection{Zero magnetic field}

Zero magnetic field corresponds to the line $q=1/2$ on the $p,q$
plane. Above we identified this line with the vertical energy axis
in Fig. \ref{levit}. Since the key ingredient of the levitation
scenario is that all the states on this axis are localized, we
start with studying localization properties along the line
$q=1/2$.
\begin{figure}[t]\vspace{-0.4cm}
\centerline{\includegraphics[width=90mm,angle=0,clip]{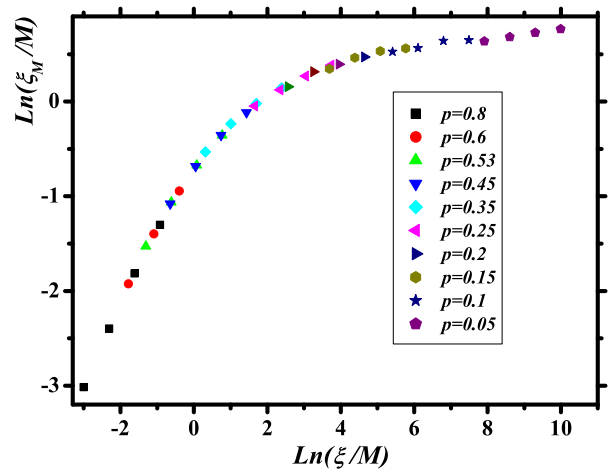}}
\caption{(Color online) A fit of $\xi_M/M$ data points for the p-q
model at $q=1/2$ to a one-parameter scaling form:
$\ln\left(\xi_M/M\right)$ vs. $\ln\left(\xi/M\right)$.}
 \label{logscale}
\end{figure}

\begin{figure}[b]\vspace{-0.4cm}
\centerline{\includegraphics[width=90mm,angle=0,clip]{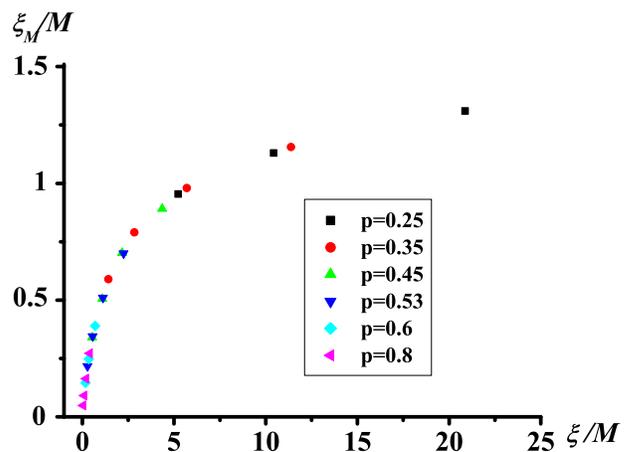}}
\caption{(Color online) One parameter scaling function,
$\xi_M/M=f(\xi/M)$, is plotted from Fig. \ref{logscale}.}
 \label{scalefig}
\end{figure}

\begin{figure}[t] \vspace{-0.4cm}
\centerline{\includegraphics[width=100mm,angle=0,clip]{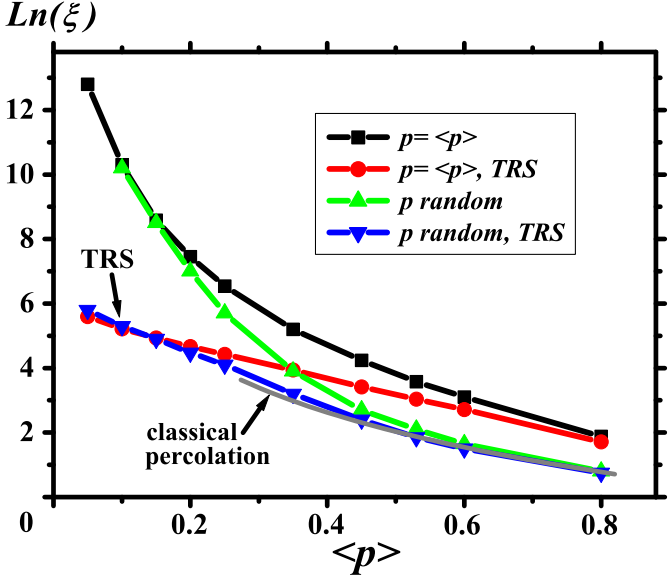}}
\caption{(Color online) Localization radius of the p-q model in a
zero magnetic field ($q=1/2$) is plotted vs. inverse "energy",
$\langle p\rangle$ for the cases: no spread in scattering
strengths on the links, $p=\langle p\rangle$, and no TRS on the
links (black), $p=\langle p\rangle$, with TRS on the links (red).
Blue and green curves show $\ln\xi$ with and without TRS,
respectively, plotted for the strong disorder in scattering
strengths on the links: $p$ randomly assumes the values $0.01$ and
$0.99$ while the inverse "energy" is the average value, $\langle
p\rangle$. Grey curve [$\ln(\xi/\xi_0)=-\frac83\ln p$] is the
result of the percolation treatment Eq. (\ref{center}). The
constant $\xi_0$ is chosen to match the data at $\langle
p\rangle=0.8$.}
 \label{q05plot}
\end{figure}

To obtain a scaling plot, $\xi_M/M=f(\xi(p)/M)$, we analyzed 40
data points: four $M$-values, $M=16,\,32,\,64,\,128$, and ten
$p$-values,
$p=0.05,\,0.1,\,0.15,\,0.2,\,0.25,\,0.35,\,0.45,\,0.53,\,0.6,\,0.8$.
By plotting $\ln\left(\xi_M/M\right)$ vs. $-\ln M$ and shifting
points for different $p$ to fall on the same line, we get the
result shown in Fig. \ref{logscale}. It is seen that the quality
of scaling is high. The scaling function, $f$, found from the data
in Fig. \ref{logscale}, is shown in Fig. \ref{scalefig}. The
dependence, $\xi(p)$ inferred in this way is plotted in Fig.
\ref{q05plot} with the black line. Our result confirms the
expectation that the localization length in zero magnetic field
increases rapidly even in log-scale as $p$ goes to zero. Scaling
theory predicts the dependence, $\ln\xi_u\sim
(k_{\scriptscriptstyle F}l)^2$, Eq. (\ref{unitary}). Since we have
earlier identified $k_{\scriptscriptstyle F}l$ with $1/p$, we
expect the dependence, $\ln\xi(p)\sim 1/p^2$. The black curve in
Fig. \ref{q05plot} falls off with $p$ slower, and can be well
approximated with $\ln\xi(p)=0.99-4\ln p$. One possibility to
account for this discrepancy is that asymptotic $1/p^2$ behavior
is achieved at $p$ smaller that $0.05$- the minimal $p$ we
studied. Note however that the total range of change of $\xi(p)$
in the domain we studied is huge: $\xi(0.05)/\xi(0.8)\approx
5.5\times10^4$.
%

\begin{figure}[t]
\centerline{\includegraphics[width=90mm,angle=0,clip]{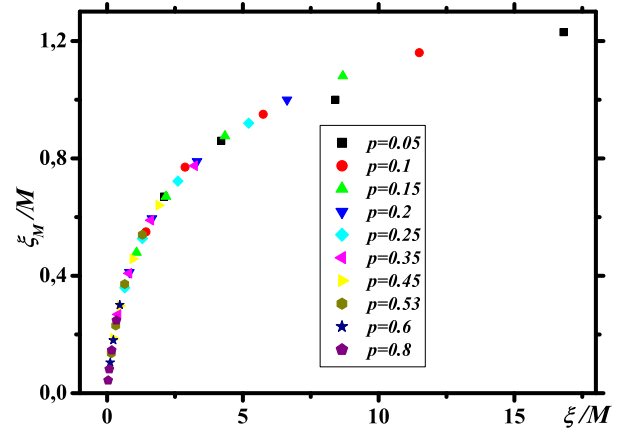}}
\caption{(Color online) A fit of $\xi_M/M$ data points for the p-q
model at $q=1/2$ with time reversal symmetry on the links to a
one-parameter scaling form $\xi_M/M=f(\xi/M)$.}
 \label{scaleTRS}
\end{figure}

Another prediction of the scaling theory is that the presence of
time reversal symmetry (TRS) the growth of localization radius
with $k_{\scriptscriptstyle F}l$ is much slower, $\ln\xi_o\sim
k_{\scriptscriptstyle F}l$, Eq. (\ref{orthogonal}). In the above
simulations we assumed that magnetic field is zero in the
"orbital" sense ($q=1/2$), but phases on the links were "unitary".
This is because the orthogonal-unitary crossover takes place at
exponentially small $(1/2-q)$. However, in order to relate closer
our calculation to the scaling theory, we ran simulations with TRS
on the links restored. This amounts to setting
$\varphi_2=-\varphi_1$, $\varphi_4=-\varphi_3$, $\phi_2=-\phi_1$,
and $\phi_4=-\phi_3$ in Eq. (\ref{Tmatrix}). The scaling function
obtained with TRS is shown in Fig. \ref{scaleTRS}, and the
corresponding $\xi(p)$ is plotted in Fig. \ref{q05plot} with the
red curve. As could be expected, in the strongly localized domain,
$p>0.6$, there is no difference between the unitary and orthogonal
cases. Upon decreasing $p$, orthogonal $\xi(p)$ indeed grows much
slower than unitary $\xi(p)$, so that $\xi(0.05)/\xi(0.8)\approx
48.5$. The fact that orthogonal $\ln\xi(p)$ extrapolates at
$p\rightarrow0$ to a finite value also suggests that diverging
behavior sets in at $p$ smaller than $0.05$.

In the above analytical treatment of the p-q model we considered
the limit of strong disorder, which is a strong spread in the
local values of $p$ with average $p=\langle p\rangle$ fixed.
Consideration was based on the understanding that strong spread in
$p$ eliminates completely the interference effects, and thus
reduces the analysis of the p-q model to the percolation problem,
which predicts much smaller localization radius,
$\xi(p)\sim1/p^{8/3}$, Eq. (\ref{center}). In order to test this
expectation, we incorporated a strong spread in $p$ into
transfer-matrix calculation. Namely, for a fixed $\langle
p\rangle$ we randomly set the  values $p=0.01$ or $p=0.99$ on each
link. Our results on $\xi(p)$ with and without TRS  are shown in
Fig. \ref{q05plot} with the blue and green lines, respectively. We
see that our expectation is confirmed within the region $p>0.5$.
In this region, the two curves with disorder are not sensitive to
universality class and are well below the curves without disorder
in $p$. Moreover, their behavior in this region is in accord with
prediction of percolation theory. Indeed, percolation theory
predicts $\xi(0.53)/\xi(0.8)=(0.8/0.53)^{8/3}\approx3$. From the
blue and green curves we get the close values, $3$ and $3.67$,
respectively. Fig. \ref{q05plot} also illustrates that for
$p<0.5$, quantum mechanics "wins" over disorder: the curves with
random $p$ merge with curves without disorder in $p$ corresponding
to their respective symmetry classes.

The above results pose an acute question: whether the boundary,
$p_c(q)$, of delocalization transitions, which was established
within the percolation treatment, and extends in the region
$p<0.5$, is preserved in fully quantum limit. A related question
is whether this boundary is sensitive to the universality class.
We address these questions below.

\subsection{The line of delocalization transitions and critical
exponent}
\begin{figure}[t]
\centerline{\includegraphics[width=85mm,angle=0,clip]{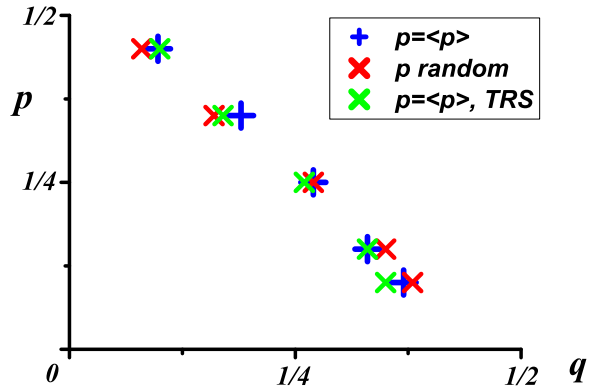}}
\caption{(Color online) Delocalization points for three different
sets of $p$: non-random $p$, random $p$ with given $\langle
p\rangle$, and non-random $p$ with the TRS.} \label{mainplot}
\end{figure}

Our main result, $p_c(q)$ boundary without TRS, is shown in Fig.
\ref{phasediag} with crosses. Five points obtained for $p=0.1$,
$0.15$, $0.25$, $0.35$, and $0.45$, essentially fall onto a
straight line, $p=1/2-q$. Deviation from this line takes place
near $p=1/2$, where the data-points match well the results of
percolation treatment (red line). This confirms our expectation
that percolation treatment of the p-q model correctly predicts
{\it position} of the quantum delocalization transition.
Certainly, the critical exponent of delocalization transition is
different: $\nu=7/3$ instead of $\nu=4/3$ for percolation. For
smaller $p$, deviation from the percolation treatment (green line)
is notable. On the other hand, as we explained above, our
percolation estimate is rather rude at small $p$. We also argued
that the true small-$p$ percolation boundary should be linear.
This linearity would mean that the constant, $\alpha$, in Eq.
(\ref{linear}) is $\alpha=1$.

More numerical results for p-q boundary are presented in
Fig.~\ref{mainplot}. For comparison, we reproduced the blue
crosses from Fig. \ref{phasediag} in Fig.~\ref{mainplot}. Red
crosses demonstrate that the boundary is {\it robust} against
strong disorder in backscattering strength, $p$. In the same way
as for zero magnetic field, the disorder in $p$ was incorporated
by randomly choosing local values of $p$ to be $0.99$ or $0.01$
while keeping $\langle p\rangle$ fixed. Coincidence of the two
boundaries indicates that delocalization transition is governed
exclusively by average $p$, but does not depend on local disorder.

The fact that in the CC model the position of the delocalized
state does not depend on the spread in local transmission
coefficients of the nodes (i.e., the spread in saddle point
heights), is obvious consequence of duality. On the other hand,
the p-q model does not possess duality, so that coincidence of the
two boundaries is by no means obvious. It can be interpreted as an
evidence that, even in magnetic field, localization properties of
the system are governed by zero-field conductance, $\sim 1/\langle
p\rangle$.
\begin{figure}[t]
\centerline{\includegraphics[width=80mm,angle=0,clip]{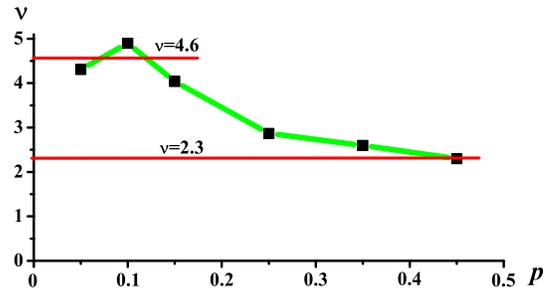}}
\caption{(Color online) Dependence of the critical exponent $\nu$
on $p$ as inferred from optimization procedure, described in the
text. As $p$ decreases from $1/2$ to $0$, $\,\nu(p)$ grows
approximately twice.}
 \label{doubl}
\end{figure}

Change of the universality class amounts to replacement
$1/\sigma_{xx}$ (unitary) by a constant (orthogonal) in the first
term of the scaling equation Eq. (\ref{pru1}). If the scaling
theory applies, this replacement should not affect the fixed
point, $\sigma_{xy}=n$. In other words, only orbital action of
magnetic field is sufficient to drive the system into quantum Hall
insulator state \cite{footnote}. We were able to check this
prediction within the p-q model. Green crosses in
Fig.~\ref{mainplot} show the position of p-q boundary with TRS.
Overall, the boundaries with and without TRS coincide. Discrepancy
at $p=0.1$ is likely due to strong fluctuations of the data at
small $p$.

Together with p-q boundary,  optimization procedure produced the
value of critical exponent, $\nu$. The dependence $\nu(p)$ is
shown in  Fig. \ref{doubl}. Horizontal lines in Fig. \ref{doubl}
are drawn to illustrate that simulations indicate apparent
doubling of the critical exponent, discussed for classical
percolation in the previous section. As follows from this
discussion, the doubling, revealed by numerics, is a consequence
of two close delocalized states. If scaling analysis could be
carried our in the immediate vicinity of a given delocalized
state, it would recover the conventional value, $\nu\approx 7/3$.
Indeed, for $p=0.45$, where two delocalized states are far apart,
optimization yields $\nu(0.45)=2.3$, while for $p=0.1$, where the
fluctuations are strong, we get $\nu(0.1)=4.9$, which is even
bigger than $2\cdot 7/3$. What is remarkable about Fig.
\ref{doubl} is that the apparent growth of $\nu$ upon decreasing
$p$ starts quite early, {\it e.g.}, for $p=0.25$ we get $\nu=2.9$.
To make sure that optimization does not distort the raw data, we
have checked the scaling manually and reproduced the largest and
smallest values of $\nu$.

\section{Triangular p-q model}

\begin{figure}[t]
\centerline{\includegraphics[width=85mm,angle=0,clip]{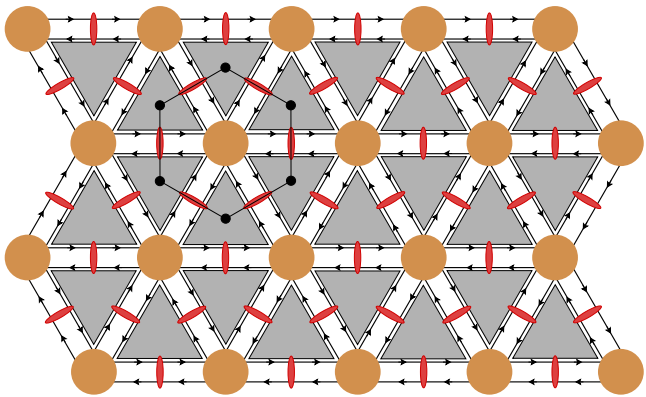}}
\caption{(Color online) Triangular version of the "square" p-q
model, Fig. \ref{pqnetwork}. Point contacts on the links stand for
the same $2\times2$ scattering matrix as in Fig. \ref{pqnetwork},
while the bend-junctions are described by a $6\times 6$ matrix Eq.
(\ref{direct}).
The centers of the forbidden regions (shaded triangles) constitute
a hexagonal lattice. }
 \label{triPQ}
\end{figure}

In the previous consideration, electron motion was restricted to
the channels between forbidden regions, $A_{n,m}$, with centers
residing on a square lattice. This consideration led us to the
phase diagram Fig. \ref{phasediag}, containing the line of
delocalization transition in the $p,q$- plane. In the present
section we will demonstrate that the same shape of the transition
line emerges when the centers of the forbidden regions constitute
a {\it hexagonal} lattice, as shown in Fig. \ref{triPQ}. This
figure illustrates that, similarly to the square p-q model,
backscattering takes place on the links. Fig. \ref{triPQ} also
illustrates that for hexagonal arrangement of forbidden regions, a
junction corresponds to the point where {\it six} such regions
come close.

By contrast to the square p-q model, a wave incident on a junction
can be scattered not into two but rather into three directions.
Namely, it can proceed forward, or get deflected by the angles
$\pm \pi/3$, Fig. \ref{triPQ}. Recall that for the square p-q
model the $4\times4$ scattering matrix of a junction Eq.
(\ref{Matrixq}) had a simple form, namely, a direct product of two
$2\times2$ matrices. Correspondingly, the $6\times6$ junction
scattering matrix in Fig. \ref{triPQ} is a direct sum of two
$3\times3$ matrices [see Eq. (\ref{direct}) below]. This is a
consequence of the fact that the two channels on a given link are
not mixed by the junction. As a result, similarly to the square
p-q model, upon switching off the backscattering on the links, the
network breaks into two decoupled fully chiral networks. While for
the square p-q model each chiral network was of CC type, here each
chiral network represents a "triangular" model introduced in Ref.
\onlinecite{Triangular}. This chiral model, illustrated in Fig.
\ref{trinet}, is called triangular because the nodes are arranged
on a triangular lattice. For convenience, we briefly review the
triangular network model Ref. \onlinecite{Triangular} in the next
subsection.

\subsection{Fully chiral triangular model}

\begin{figure}[t]
\centerline{\includegraphics[width=80mm,angle=0,clip]{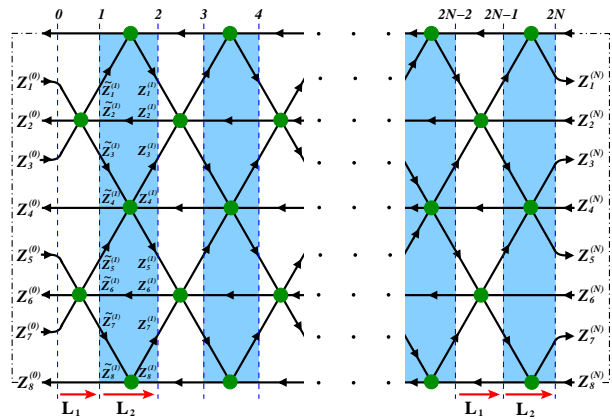}}
\caption{(Color online) A slice of triangular network of width
$M=8$ is shown. Three amplitudes on the links to left of green dot
are related to three amplitudes to the right of green dot via
matrix $\mathbf{X}$. The upper and the lower boundaries of the
slice are connected by dashed-dotted lines manifesting that the
amplitudes on these boundaries are the same by virtue of cyclic
boundary conditions. Upon passing the white stripe, the vector of
the amplitudes $\{Z_i\}$ is multiplied by the matrix
$\mathbf{L}_1$.
 Upon passing the blue stripe the vector $\{\tilde{Z}_i\}$
is multiplied by $\mathbf{L}_2$.}
 \label{trinet}
\end{figure}

\begin{figure}[b]
\centerline{\includegraphics[width=90mm,angle=0,clip]{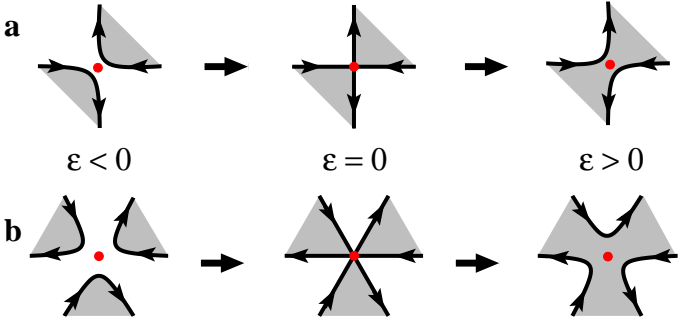}}
\caption{(Color online) Evolution of the equipotential lines: (a)
in the CC model, (b) in the chiral triangular model.}
 \label{evolution}
\end{figure}

Microscopic picture underlying CC and triangular chiral models is
illustrated in Fig. \ref{evolution}. Fully chiral motion
represents drift of the Larmour circle along equipotential lines
of smooth random potential. In CC model equipotentials meet
pairwise at the saddle points. As the energy, $\varepsilon$,
passes through zero, the geometry of equipotentials evolves from
reflection to transmission, as shown in Fig. \ref{evolution}a. The
scattering matrix describing this evolution near $\varepsilon=0$
has the form \cite{CC}
\begin{eqnarray}
\label{CCsc} &&S_{cc}(\varepsilon)=\left(\begin{array}{cc}
\,\frac1{\sqrt{2}}+\varepsilon&\frac1{\sqrt{2}}-\varepsilon\\
\,\\
-\frac1{\sqrt{2}}+\varepsilon&\frac1{\sqrt{2}}+\varepsilon\\
\end{array}\right).
\end{eqnarray}
At $\varepsilon=0$ the power reflection and transmission
coefficients are both equal to $1/2$.

In triangular model it is assumed that smooth potential has
$120^\circ$ rotational symmetry. As a result, equipotentials meet
in the groups of three. As $\varepsilon$ is swept through zero,
they evolve from reflection (shaded regions disconnected) to
transmission (shaded regions fully connected), see Fig.
\ref{evolution}b. Near $\varepsilon=0$, probabilities of
scattering to the left, to the right and forward, are all finite.
Obviously, at $\varepsilon=0$ probabilities of the left- and
right- scattering are equal to each other. In Ref.
\onlinecite{Triangular} it was demonstrated that these
probabilities are equal to $4/9$. Correspondingly, the probability
of the forward scattering is $1/9$. At small but finite
$\varepsilon$ the form of scattering matrix is dictated by
$\varepsilon\rightarrow-\varepsilon$ duality and flux
conservation. Up to $\varepsilon^2$ terms it is given by
\begin{eqnarray}
\label{sc}
&&S_{\vartriangle}(\varepsilon)=\left(\begin{array}{ccc}
\,\frac23(1+\varepsilon)&-\frac13&\,\frac23(1-\varepsilon)\\
\,\\
\,\frac23(1-\varepsilon)&\,\frac23(1+\varepsilon)&-\frac13\\
\,\\
-\frac13&\,\frac23(1-\varepsilon)&\,\frac23(1+\varepsilon)
\end{array}\right).
\end{eqnarray}
Recall that in the limit of strong disorder, when the black
regions in Fig. \ref{evolution}a are either connected or fully
disconnected, the CC model reduces to the {\it bond} percolation
problem on a square lattice. Correspondingly, the strong-disorder
limit of the chiral triangular model is the {\it site} percolation
on a triangular lattice \cite{Triangular}. In the strong disorder
limit, $\varepsilon=0$ corresponds to the equal portion of present
and absent sites. In accord to well-known result, this site
percolation problem possesses a property of self-duality
\cite{SykesEssam}.
\begin{figure}[b]\vspace{-0.4cm}
\centerline{\includegraphics[width=80mm,angle=0,clip]{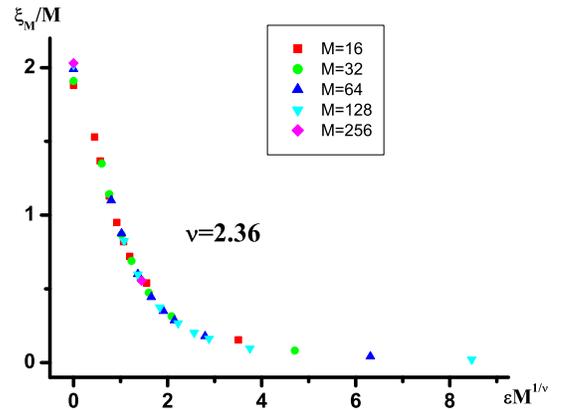}}
\caption{(Color online) Numerical results for fully chiral
triangular network model. A fit of data to a one-parameter scaling
form, $\xi_M/M=f(\varepsilon M^{1/\nu})$ yields the critical
exponent, $\nu\approx 2.36$.}
 \label{tricrexp}
\end{figure}

In Ref. \onlinecite{Triangular} the critical exponent of the
triangular model Fig. \ref{trinet} was inferred from the real
space renormalization group analysis. The result, $\nu \approx
2.3\div2.76$, agrees with simulations of the CC model reported in
the literature. Here we present the result of numerical
simulations of the triangular model. Simulations use the slice
transfer matrix
\begin{eqnarray}
\label{tm}
\mathbf{T}=\prod_{n=N-1}^{0}\mathbf{L}_2\mathbf{P}_2^{(n)}\mathbf{L}_1
\mathbf{P}_1^{(n)},
\end{eqnarray}
Operators $\mathbf{L}_1$ and $\mathbf{L}_2$ act in "white" and
"blue" stripes in Fig. \ref{trinet}, respectively. The operator
$\mathbf{L}_1$ performs transformation of the vector of
amplitudes, $\{Z_i^{(n)}\}$, into $\{\tilde{Z}_i^{(n+1)}\}$, while
$\mathbf{L}_2$ performs transformation of $\{\tilde{Z}_i^{(n)}\}$
into $\{Z_i^{(n)}\}$, see Fig. \ref{trinet}. For a particular
stripe width, $M=8$, the matrix forms of $\mathbf{L}_1$ and
$\mathbf{L}_2$ are the following
\begin{eqnarray}
\mathbf{L}_1 =\left( \begin{array}{cccc|cccc}
x_{11}& x_{12}& x_{13}& 0 &\cdot &\cdot &\cdot &\cdot \\
x_{21}& x_{22}& x_{23}& 0 &\cdot &\cdot &\cdot &\cdot \\
x_{31}& x_{32}& x_{33}& 0 &\cdot &\cdot &\cdot &\cdot \\
0&0 &0 &1 &\cdot &\cdot &\cdot &\cdot  \\
\hline
\cdot &\cdot &\cdot &\cdot &x_{11}& x_{12}& x_{13}& 0 \\
\cdot &\cdot &\cdot &\cdot &x_{21}& x_{22}& x_{23}& 0 \\
\cdot &\cdot &\cdot &\cdot &x_{31}& x_{32}& x_{33}& 0 \\
\cdot &\cdot &\cdot &\cdot &0&0 &0 &1
\end{array} \right),
\end{eqnarray}
\begin{eqnarray}
\mathbf{L}_2 =\left( \begin{array}{cc|cccc|cc}
 x_{33}&0 &\cdot &\cdot &\cdot &\cdot & x_{31}& x_{32} \\
0&1&\cdot &\cdot &\cdot &\cdot &0&0 \\
\hline
\cdot &\cdot & x_{11}& x_{12}& x_{13}& 0 &\cdot &\cdot \\
\cdot &\cdot & x_{21}& x_{22}& x_{23}& 0 &\cdot &\cdot \\
\cdot &\cdot & x_{31}& x_{32}& x_{33}& 0 &\cdot &\cdot \\
\cdot &\cdot &0 &0 &0 &1 &\cdot &\cdot   \\
\hline
x_{13}&0 &\cdot &\cdot &\cdot &\cdot & x_{11}& x_{12} \\
x_{23}&0 &\cdot &\cdot &\cdot &\cdot& x_{21}& x_{22}
\end{array} \right),
\end{eqnarray}
with dots standing for zeroes. In the above relations,
$\{x_{ij}\}$ form a $3\times3$ matrix, $\mathbf{X}$, which is the
node transfer matrix corresponding to the scattering matrix Eq.
(\ref{sc}), and up to $\varepsilon^2$ terms has the matrix form
\begin{eqnarray}
\label{ns2} \mathbf{X}= \left(\begin{array}{ccc}
2(1+\varepsilon)&-2(1+\varepsilon)&1\\
\,\\
2(1+\varepsilon)&-3&2(1-\varepsilon)\\
\,\\
1&-2(1-\varepsilon)&2(1-\varepsilon)
\end{array}\right).
\end{eqnarray}
Specific form of $\mathbf{L}_2$ accounts for the cyclic boundary
conditions in the vertical direction. Matrices
\begin{eqnarray}
\mathbf{P}_1^{(n)}=\text{diag}\{e^{i\varphi_1^{(n)}},
\cdots,e^{i\varphi_8^{(n)}}\}
\end{eqnarray}
and
\begin{eqnarray}
\mathbf{P}_2^{(n)}=\text{diag}\{e^{i\psi_1^{(n)}},
\cdots,e^{i\psi_8^{(n)}}\}.
\end{eqnarray}
account for the random phases on the links.

Similarly to CC model, the critical properties of the triangular
model were inferred from the scaling analysis of the Lyapunov
exponents. The scaling plot is shown in Fig. \ref{tricrexp}. A
high quality scaling was achieved for $\nu=2.36$, which is in a
good agreement with simulations of the CC model.

\subsection{Numerical results for the triangular p-q model}
\begin{figure}[t]\vspace{-0.4cm}
\centerline{\includegraphics[width=90mm,angle=0,clip]{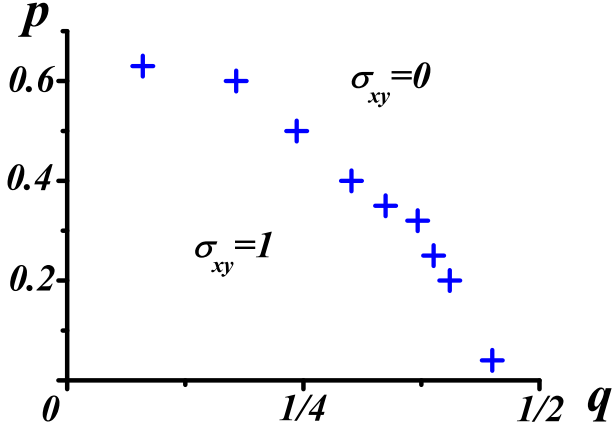}}
\caption{(Color online) Numerical results for the line of
delocalization transitions in triangular p-q model. The point
$q=0$ corresponds to $p=p_c=0.6527$- threshold of bond percolation
on hexagonal lattice.}
 \label{tridelocal}
\end{figure}

In the CC model, which is based on the picture of equipotentials,
the scattering matrix at a node, $S_{CC}$, is a function of
energy, $\varepsilon$. By contrast, in the p-q model, describing
low magnetic fields, the scattering matrix at the node depends on
"magnetic field", $1/2-q$, while the energy dependence enters via
the backscattering probability, $p$. Still, the structure of the
scattering matrices of the CC model and the p-q model are the
same. It is also the case for the triangular p-q model, for which
we choose the following form of the $q$- dependent scattering
matrix
\begin{equation}
\label{qsm}
S=\frac{1}{q^2-q+1}
\left(\begin{array}{ccc} 1-q &q(q-1)
&q\\
\,\\
q &1-q &q(q-1)\\
\,\\
q(q-1) &q &1-q
\end{array}\right).
\end{equation}
This matrix is unitary for all $0\leq q\leq 1$. It is critical at
$q=1/2$. Indeed, as follows from Eq. (\ref{qsm}), the ratio of
probabilities of scattering to the left and to the right is
$q/(1-q)$, so that at $q=1/2$ these probabilities are equal. Their
values are $4/9$ in agreement with Eq. (\ref{sc}). With this
parametrization, the $6\times6$ junction scattering matrix in Fig.
\ref{triPQ} acquires the form
\begin{equation}
\label{direct}
{\scriptstyle \frac{1}{q^2-q+1}}\! \left(\!
\begin{array}{cccccc}
{\scriptstyle 0} &\!\!{\scriptstyle 1-q} &\!\!{\scriptstyle 0} &\!\!{\scriptstyle q(q-1)} &\!\!{\scriptstyle 0} &\!\!{\scriptstyle q}\\
{\scriptstyle 1-q} &\!\!{\scriptstyle 0} &\!\!{\scriptstyle q(q-1)} &\!\!{\scriptstyle 0} &\!\!{\scriptstyle q} &\!\!{\scriptstyle 0}\\
{\scriptstyle 0} &\!\!{\scriptstyle q} &\!\!{\scriptstyle 0} &\!\!{\scriptstyle 1-q} &\!\!{\scriptstyle 0} &\!\!{\scriptstyle q(q-1)}\\
{\scriptstyle q} &\!\!{\scriptstyle 0} &\!\!{\scriptstyle 1-q} &\!\!{\scriptstyle 0} &\!\!{\scriptstyle q(q-1)} &\!\!{\scriptstyle 0}\\
{\scriptstyle 0} &\!\!{\scriptstyle q(q-1)} &\!\!{\scriptstyle 0} &\!\!{\scriptstyle q} &\!\!{\scriptstyle 0} &\!\!{\scriptstyle 1-q}\\
{\scriptstyle q(q-1)} &\!\!{\scriptstyle 0} &\!\!{\scriptstyle q}
&\!\!{\scriptstyle 0} &\!\!{\scriptstyle 1-q} &\!\!{\scriptstyle
0}
\end{array}\!\!\!\right).
\end{equation}

Since backscattering in triangular p-q model, Fig. \ref{trinet},
takes place on the links, it is described by the same $2\times2$
matrix Eq. (\ref{Matrixp}) as in the square p-q model. Thus, by
analogy to Eq. (\ref{Tmatrix}), the $T$- matrix of the triangular
p-q model has the form
\begin{widetext}
\begin{equation}
\label{TTrimat}  \hat{\text{\large T}}=\! \left(\!\!
\begin{array}{cccccc}
\tilde{x}_{11}&0&0&\tilde{x}_{12}&\tilde{x}_{13}&0\\
0&x_{11}&x_{12}&0&0&x_{13}\\
0&x_{21}&x_{22}&0&0&x_{23}\\
\tilde{x}_{21}&0&0&\tilde{x}_{22}&\tilde{x}_{23}&0\\
\tilde{x}_{31}&0&0&\tilde{x}_{32}&\tilde{x}_{33}&0\\
0&x_{31}&x_{32}&0&0&x_{33}
\end{array}\right)
 \hat{\text{\large D}}_1
\left(\!\!
\begin{array}{cccccc}
{\scriptstyle\frac1{\sqrt{1-p}}}&{\scriptstyle \sqrt{\frac p{1-p}}}&0&0&0&0\\
{\scriptstyle \sqrt{\frac p{1-p}}}
&{\scriptstyle\frac1{\sqrt{1-p}}}&0&0&0&0\\ 0&0&
{\scriptstyle\frac1{\sqrt{1-p}}} &{\scriptstyle \sqrt{\frac p{1-p}}}&0&0\\
0&0&{\scriptstyle \sqrt{\frac p{1-p}}}
&{\scriptstyle\frac1{\sqrt{1-p}}}&0&0\\
0&0&0&0&{\scriptstyle\frac1{\sqrt{1-p}}}&{\scriptstyle \sqrt{\frac
p{1-p}}}\\
0&0&0&0&{\scriptstyle \sqrt{\frac p{1-p}}}
&{\scriptstyle\frac1{\sqrt{1-p}}}
\end{array}\!\!\right)
 \hat{\text{\large D}}_2,
\end{equation}
\end{widetext}
where the diagonal matrices
\begin{eqnarray}
\hat{\text{\large D}}_1=\text{diag}\{e^{i\varphi_1},
\cdots,e^{i\varphi_6}\}
\end{eqnarray}
and
\begin{eqnarray}
\hat{\text{\large D}}_2=\text{diag}\{e^{i\psi_1},
\cdots,e^{i\psi_6}\}
\end{eqnarray}
account for the random phases on the links. The matrix,
$\{x_{ij}\}=\text{\large X}$, has the explicit $q$- dependence
\begin{equation}
\label{Xq}
\text{\large X}(q)= \left(\begin{array}{ccc} \frac1q &-\frac1q
&1\\
\,\\
\frac1q &\frac{q^2-q+1}{q(q-1)} &\frac1{q-1}\\
\,\\
1 &\frac1{1-q} &\frac1{1-q}
\end{array}\right),
\end{equation}
and $\{\tilde{x}_{ij}\}=\tilde{\text{\large X}}$ are given by the
relation
\begin{equation}
\label{hexbond} \tilde{\text{\large X}}(q)=\text{\large
X}^{-1}(1-q).
\end{equation}

In the square p-q model, two prominent points on the $p,q$ plane
were $(\frac12,0)$ and $(0,\frac12)$. They corresponded to the
bond percolation over p- and q- bonds, respectively. Similarly, in
the triangular p-q model, the point $(0,\frac12)$ is
distinguished. At this point, due to the absence of p- bonds, two
decoupled q- subnetworks undergo the site percolation. In
triangular p-q model, the counterpart of the p- bond percolation
at $(\frac12,0)$ of the square p-q model is a p- bond percolation
on a {\it hexagonal} lattice. This is because the centers of
forbidden regions in Fig. \ref{triPQ} constitute a hexagonal
lattice. Thus, the second distinguished point for triangular p-q
model should be $(p_c,0)$, where
\begin{equation}
\label{hexbond} p_c=1-2\sin\left(\frac\pi{18}\right)=0.6527
\end{equation}
is the threshold of the bond percolation on the honeycomb lattice
\cite{SykesEssam}. Our numerical simulations based on the transfer
matrix Eq. (\ref{TTrimat}) confirm this expectation. The points of
the delocalization transition, shown in Fig. \ref{tridelocal}
follow the line which smoothly connects the point $(0.6527, 0)$
and $(0,\frac12)$. The general shape of the line is quite similar
to the above results, Fig. \ref{phasediag}, for the square p-q
model. In particular, in the domain of vanishing magnetic fields,
$(1/2-q)\ll1$, the transition boundary is again linear, in
agreement with prediction Eq. (\ref{positions}) of the scaling
theory.

In conclusion of this section we would like to make the following
remark. Simplification of the node structure in the square p-q
model was achieved by choosing the matrix Eq. (\ref{Matrixq})
which captures orbital action of magnetic field, but does not
allow forward- and backward scattering, which, instead, takes
place on the links. Similarly, in the triangular p-q model, the
junction matrix Eq. (\ref{direct}) restricts the scattering
options for incident electrons. Namely, out of six possibilities,
the electron can be scattered only into three channels. Again,
similarly to the square p-q model, it can access the three other
channels upon backscattering on the links.

\section{Strongly localized region: implications for inelastic transport}
\begin{figure}[t]\vspace{-0.4cm}
\centerline{\includegraphics[width=90mm,angle=0,clip]{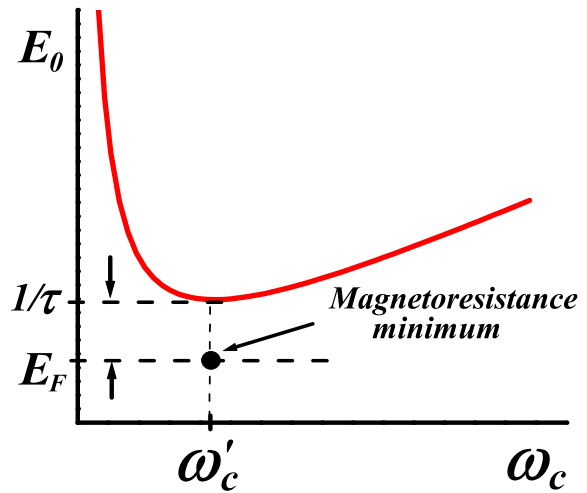}}
\caption{(Color online) Levitation plot for $n=0$; same as lower
curve in Fig. \ref{levit}. For electron densities corresponding to
$E_F$ slightly below the minimum, inelastic magnetoresistance has
a deep minimum near magnetic field, $\omega_c=\omega_c^\prime$.}
 \label{magres}
\end{figure}

The presence of a minimum in the line, $E_n(\omega_c)$, of the
delocalization transitions, Fig. \ref{levit}, manifests itself in
specific behavior of low-temperature magnetoresistance,
$\rho_{xx}(\omega_c)$. For high enough electron densities,
$\rho_{xx}(\omega_c)$ is temperature independent at two (low and
high) distinct values of magnetic field. This $T$- independence is
a signature of the delocalization transition. Different
experimental groups have studied details of the behavior of
$\rho_{xx}(\omega_c,T)$ near the high-field transition
\cite{shahar95,shahar96,dunford00,Ponomarenko07}, low-field
transition \cite{shahar97,shahar'97}, or both transitions
\cite{Exp6,hilke97,hilke98,hilke00,murzin02,yasin02}. For Fermi
level position below the minimum in Fig. \ref{levit}, the system
remains localized upon increasing $\omega_c$, see Fig.
\ref{magres}. However, even in this localized regime, proximity of
$\omega_c$ to the position of minimum, $\omega_c^\prime$, (Fig.
\ref{magres}) should manifest itself as a precursor of
delocalization in the {\it inelastic} transport. Such a precursor
of delocalization has actually been observed in the early papers
\cite{Jiang92, Koch92} in the form of a minimum in
$\rho_{xx}(\omega_c)$, in the regime of the variable-range
hopping. This minimum reflects the increase of localization
radius, $\xi(\omega_c)$, near $\omega_c=\omega_c^\prime$. The
lower is the temperature, the deeper is minimum, as follows from
the Mott's law
\begin{equation}
\label{Mott} \ln\left[\rho_{xx}(\omega_c)\right]\propto\frac
1{\left[\xi^2(\omega_c)T\right]^{1/3}}.
\end{equation}
Theoretical studies of variable-range-hopping magnetoresistance
pertained to deeply localized regime, where interference effects
in a single hopping act constituted a small but singular in
$\omega_c$ correction to the tunneling probability
\cite{Shklovskii85, Shklovskii86}. The correction is small because
the tunneling probability is greatly reduced if electron
under-barrier trajectory deviates from the straight line.
According to Refs.~\onlinecite{Shklovskii85},
\onlinecite{Shklovskii86}, interference responsible for
magnetoresistance occurs between different virtual
forward-tunneling trajectories. The other early theory Ref.
\onlinecite{AAK82} of negative hopping magnetoresistance was based
on the following reasoning. Weak magnetic field, by changing the
universality class and thus inducing delocalization of states with
high energies, $\gtrsim1/\tau$, causes some growth of $\xi$ for
the states in the deep tail. Note that both theories, as well as
later theory Ref. \onlinecite{Glazman}, were based on the {\it
phase} rather than orbital action of magnetic field.
\begin{figure}[t]\vspace{-0.4cm}
\centerline{\includegraphics[width=100mm,angle=0,clip]{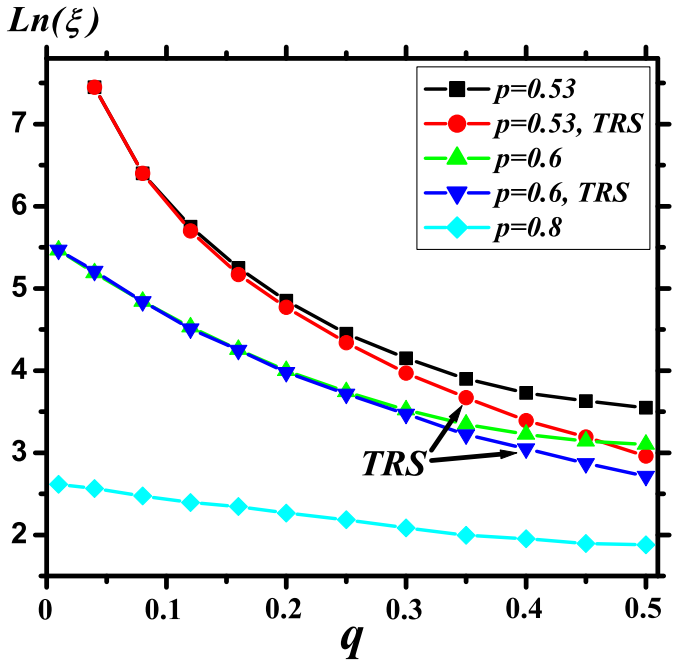}}
\caption{(Color online) Localization radius in the strongly
localized region $p>1/2$ is plotted vs. magnetic field. The point
$q=0$ represents the field, $\omega_c=\omega^\prime_c$ in Fig.
\ref{magres}. Red and black curves for $p=0.53$, and blue and
green curves for $p=0.6$ corresponding to TRS and no TRS,
respectively, merge upon increasing magnetic field. Light blue
curve corresponds to $p=0.8$. There is no difference between TRS
and no TRS for this deeply localized energy.}
 \label{insulating}
\end{figure}

By now there is no theory describing the behavior of localization
radius, $\xi(E_{\scriptscriptstyle F}, \omega_c)$, in the domain
of intermediate fields $\omega_c\sim\omega_c^\prime\sim1/\tau$ and
electron densities ( Fermi energies, $ E_{\scriptscriptstyle F}$)
at which electron states are localized but not strongly, so that
$E_{\scriptscriptstyle F}\tau\sim k_{\scriptscriptstyle F}l\sim1$.
On the other hand, the p-q model offers a unique possibility to
study localization length in this domain. To this end, we studied
numerically $\xi(p,q)$ in the region $p>1/2$ on the $p,q$ plane.
The dependence $\xi(p,q)$ can be translated into the
$\xi(E_{\scriptscriptstyle F}, \omega_c)$ dependence. We find that
the phase action of magnetic field {\it does} lead to a certain
increase in $\xi$ at low fields. However, further increase of
$\omega_c$, when the orbital action sets in, causes a much
stronger delocalization effect.

Our numerical results for $\xi(p,q)$ are presented in
Fig.~\ref{insulating}, for the values $p=0.53,\,0.6,$ and $0.8$.
In order to illuminate the role of the mechanism
Refs.~\onlinecite{Shklovskii85}, \onlinecite{Shklovskii86},
simulations were performed with and without TRS on the links. The
values of $\xi(p,q)$ were inferred from the scaling analysis of
$\xi_M/M$, similarly to the case of zero magnetic field. For
$p=0.8$, the difference between TRS and no-TRS is negligibly
small. Still the interplay of orbital effect and interference
leads to enhancement of $\xi$ from weak, $q=1/2$, to strong,
$q=0$, magnetic fields, by a factor of $2$. From Eq. (\ref{Mott})
it follows that the corresponding drop-off of log-resistance is
$2^{2/3}$, and $\ln(\rho)$ is approximately linear in $(1/2-q)$.
This allows a comparison with the experimental data of Ref.
\onlinecite{Jiang92}, where giant negative hopping
magnetoresistance of a degenerate 2D electron gas was reported.
For experimental value of $\xi=250\, \AA $ the low-field
log-resistance at temperature, $T=0.3K$, was $8.7$. The net
drop-off of log-resistance observed was $2.3$. This corresponds to
the increase of $\xi$ by a factor of $1.5$. However, at $p=0.8$
our numerics suggests a linear change of $\xi$ with $q$, while
experimentally it changes slower. The possible origin of this
discrepancy lies in the fact that $(1/2-q)$ is proportional to
$\omega_c$ only at low fields. In fact, at higher fields,
$(1/2-q)$ changes slower than $\omega_c$.

As $p$ decreases towards $1/2$, both phase and orbital mechanisms
of magnetoresistance become stronger. As follows from Fig.
\ref{insulating},  the (zero-field) $\xi$ increases from TRS to no
TRS by a factor $1.47$ and $1.82$ for $p=0.6$ and $p=0.53$,
respectively. This increase is a quantitative measure of the phase
mechanism. It should result in drop-off of the log-resistance by
the factors $1.3$ and $1.5$, which corresponds to the resistance
drop by several times. The orbital mechanism causes a
significantly bigger reduction of log-resistance. The $p=0.6$ and
$p=0.53$ curves in Fig. \ref{insulating} indicate the decrease in
$\xi$ between $q=0$ and $q=0.5$ by factors $11$ and $60$,
respectively. Both effects are way too strong in comparison with
all the experimental data on hopping transport reported in the
literature. This is because $p=0.6$ and $p=0.53$ correspond to
very large values of localization radius, e.g., $\xi=3000 \AA$ in
experimental conditions of Ref. \onlinecite{Jiang92}. Temperatures
required to observe hopping transport with such large $\xi$ are
unreasonably low.

\section{Conclusion}

%
%

It is expected on general grounds that, as disordered system
undergoes a quantum Hall transition, its behavior is universal,
i.e., in the vicinity of transition this behavior does not depend
on the type of disorder. For quantum Hall transition in a strong
magnetic field numerics provides a compelling evidence for such a
universality. For example, critical exponent derived from the CC
model, based on the picture of a smooth disorder, and from the
simulations \cite{Huckestein90} for the point-like scatterers are
the same.

Scaling scenario of Ref. \onlinecite{AALR} suggests that
regardless of the type of the disorder, a single characteristics
of the system, the Drude conductance $\sigma$, determines the size
of the sample, $\xi_o$ or $\xi_u$, depending on presence or
absence of TRS, at which the system becomes an insulator. By
virtue of scaling scenario, electron moving on the links of the
network and scattered at the nodes gets localized, due to
interference effects, in the same way as realistic electron moving
on a plane and scattered by random impurities. Correspondence
between two systems is established by Eq. (\ref{mfp}).

Pruisken's theory suggests that, in magnetic field, only two
characteristics, the components $\sigma_{xx}$ and $\sigma_{xy}$ of
the Drude conductivity tensor determine what value of quantized
Hall conductivity the system will acquire at large scales, when
the interference effects suppress the diagonal conductivity.

Basing on insensitivity to the character of disorder, we expect
that weakly chiral network of junctions and point contacts
represents electron gas in a non-quantizing magnetic field. To
establish the correspondence we need to express the values of
$\sigma_{xx}$  and $\sigma_{xy}$ of our network in terms of
parameters $p$ and $q$ in the Drude regime, i.e., the regime where
the interference can be neglected. Above we have already
identified $(k_{\scriptscriptstyle F} l)^{-1}$ with parameter $p$
of the p-q model. Here we elaborate on this relation.


Our main idea of modelling a competition between
interference-induced localization and magnetic-field-induced
orbital curving is to confine the orbital action to a set of
compact spatial regions-junctions, with asymmetry of scattering to
the left and to the right proportional to magnetic field. In zero
field, these junctions represent strong scatterers of electrons.
These strong scatterers come in addition to weak "intrinsic"
scatterers, that limit the mean free path, $l$, of electron gas.
If strong scatterers are sparse, they will not affect the
transport of electron gas at all. On the other hand, if they are
dense, then the mobility will be limited exclusively by scattering
off these strong scatterers. This suggests that the distance
between the scatterers should be chosen of the order of $l$.
Therefore, as we replace realistic electron motion by a motion
along the links of a network, the dimensionless lattice constant
should be chosen as $\sim k_{\scriptscriptstyle F}l$.

In our p-q network Fig. \ref{pqnetwork}, in addition to junctions
(strong scatterers) there are point contacts on the links. The
role that these point contacts play is the following. Our
junctions Fig. \ref{pqnetwork} do not provide {\it any}
backscattering. More precisely, as can be seen from Fig.
\ref{pqnetwork}, an electron starting along a given link, say, to
the right, after several scatterings off the junctions, will never
return to the starting point from the left. Thus, the junctions
alone cannot model the interference effects in realistic electron
gas, where the probability of such a return is $\sim
1/(k_{\scriptscriptstyle F}l)$. It is the point contacts that
provide possibility of backscattering in the p-q network, and
parameter $p$ is chosen $1/(k_{\scriptscriptstyle F}l)$ in order
to model the realistic return probability.

From the above reasoning, the diagonal conductivity of the p-q
model is $\sigma_{xx}\sim1/p$. In a finite magnetic field,
estimate for $\sigma_{xy}$ can be found from Eq. (\ref{HRq}):
\begin{equation}\label{Bv}
\sigma_{xy}=R_H\sigma_{xx}\sim\left(\frac12-q\right)\sigma_{xx}\sim
\frac{1/2-q}{p}.
\end{equation}
Note now, that the boundary of delocalization transition
established in the present paper, is $\alpha(1/2-q)=p$ with
$\alpha\sim1$. Then we conclude from Eq. (\ref{Bv}) that
transition occurs when the "Boltzmann" value of $\sigma_{xy}$ is
$\sim 1$. On the other hand, levitation scenario is based on the
conjecture that the classical $\sigma_{xy}=1$ does not get
renormalized upon increasing the sample size, see Eqs.
(\ref{pru1}), (\ref{pru2}). Consistency between delocalization
condition within p-q model and within scaling theory
\cite{Khmelnitskii} can be viewed as evidence that the p-q model
adequately captures microscopic physics behind the levitation
scenario.

There is a fundamental reason why the picture of the weak-field
quantum Hall transition is much more complex than the picture of
the strong-field transition. Namely, for strong-field transition
there is an {\it exact duality} of electron states above and below
critical energy. By contrast, there is no such inherent duality in
the weak-field transition. This, in particular, does not allow to
employ the quantum real-space renormalization group approach
\cite{Triangular,aram,arovas,cain,cain03,reviews} to describe this
transition analytically.

Duality with respect to the center, $\varepsilon=0$, of the Landau
level in the CC model insures that, with a strong spread in the
saddle-point heights, the scaling region narrows, but delocalized
state remains at $\varepsilon=0$. A remarkable outcome of our
numerics is that the same property holds for the weak-field
transition: we have verified that, upon introducing a strong
spread in the local values of the backscattering strength, $p$,
but keeping $\langle p \rangle$ fixed, does not effect the
position,  $p_c(q)$, of the transition point. On the other hand,
upon increasing the spread in the local values of $p$,
interference effects become progressively less relevant. This
allowed us to uncover a transparent {\em classical} picture of the
low-field quantum Hall transition, see Figs. \ref{trajectories},
\ref{diffph}, which is a counterpart of the percolation picture of
the strong-field transition \cite{Kazarinov,Iordansky,Trugman}.
The fact that levitation emerges both within the p-q model and
from the scaling equations Eqs. (\ref{pru1}), (\ref{pru2}), still
does not mean that the p-q model offers a microscopic support for
the Pruisken theory. After all, phenomenon of levitation follows
from a general argument that delocalized states have nowhere to go
but up in the vanishing magnetic field. In fact, scaling equations
Eqs. (\ref{pru1}), (\ref{pru2}) suggest a stronger message:
namely, positions of delocalized states do not change if magnetic
field does not exercise any  "phase" action. Neglecting the phase
action corresponds to replacement of the first term in Eq.
(\ref{pru1}) by a constant. We emphasize that the p-q model
supports this prediction of the Pruisken theory. This is reflected
in the coincidence (within accuracy of our numerics) of the p-q
boundary with and without TRS.

In closing, we list the questions which were previously addressed
in the frame of fully chiral CC model and are also pertinent to
the weakly chiral p-q model:

({\it i}). Two-channel network models of CC type were studied in
Refs. \onlinecite{Kagalovsky95,
kagalovsky97,LeeChalker94,LeeChalkerKo94,wen94,kagalovsky99,gruzberg99}
in connection with spin-orbit induced splitting of quantum Hall
transition \cite{LeeChalker94,LeeChalkerKo94}, disorder-induced
"attraction" of delocalized states from different Landau levels
\cite{wen94,Kagalovsky95, kagalovsky97} and quantum Hall
transition in disordered superconductors
\cite{kagalovsky99,gruzberg99}. All these considerations are based
on two {\it co-propagating} channels on each link. In p-q model,
the two channels are {\it counter-propagating} and spinless. It
would be interesting to incorporate spin-orbit coupling into the
p-q model for the following reason. In zero magnetic field and in
the presence of spin-orbit coupling there is a critical energy
above which electron states are delocalized \cite{SO,Asada04}. On
the other hand, in strong magnetic fields, spin-orbit coupling
splits discrete delocalized states
\cite{LeeChalker94,LeeChalkerKo94}. Upon decreasing the magnetic
field, both delocalized states are most likely to head towards
zero-field metal-insulator transition point. Microscopic model
describing this scenario must include orbital action of magnetic
field, spin-orbit coupling, and interference effects. This can be
accomplished upon incorporating spin degree of freedom into the
p-q model. A possible application of the above physics is graphene
in a weak magnetic filed \cite{Furusaki}. In the latter case, the
inter-valley scattering plays the role of spin-orbit coupling.

({\it ii}). In three dimensional layered system in zero magnetic
field, arbitrarily weak coupling between the layers delocalizes
electron states above a certain critical energy. In strong
magnetic field, a weak interlayer coupling smear discrete
delocalized state into metallic band \cite{Dohmen}. Matching of
these two scenarios takes place in weak magnetic fields. The p-q
model based description can be employed in this domain.

({\it iii}). Interaction-induced dephasing \cite{wang00} is
crucial for experimental observability of levitation, since the
transition is smeared when localization radius exceeds the
dephasing length. The issue closely related to dephasing is a
peculiar behavior of the non-diagonal resistivity, $\rho_{xy}$, in
the vicinity of the high-field transition
\cite{dykhne94,ruzin95,shimshoni97,pryadko,pryadko00,Zulicke,Shimshoni04,apalkov03}.
The question about behavior of $\rho_{xy}$ near the low-field
transition point can be asked and addressed within "incoherent"
p-q model. In addition, the question about interplay of disorder
and interaction previously studied for high-field transition
\cite{cobden99,mahida01,Faulhaber05,yang95,huckestein99,yang01,wang02},
can be redirected to the low-field transition.

({\it iv}). It is known that there is an intimate relation between
delocalization transition in CC model and critical behavior of
superspin chains
\cite{DHLee94,Zirnbauer94,Zirnbauer97,Kim96,Gruzberg97,Marston99}.
It would be interesting to investigate wether the p-q model
corresponds to any spin model. It is also known that CC model is
related to the Dirac Hamiltonian with disorder
\cite{Fisher94,HoChalker}. Recall that at degeneracy point, $p=0$,
$q=1/2$, the p-q model falls into two independent CC models.
Modification of the mixing of two counter-propagating channels on
the links transforms the p-q model with $q=1/2$ to the model
studied in Ref. \onlinecite{Bocquet}. This modification leads to
criticality which was studied within the sigma-model approach
\cite{Gade1, Gade2}. It would be interesting to investigate such a
modification of the p-q model away from the degeneracy point. More
specifically, the case $q\neq 1/2$, in our p-q model, differs from
the model Ref. \onlinecite{Bocquet} primarily in the scattering
matrix of junctions. Our junction matrix Eq.~(\ref{Matrixq})
describes scattering, say, to the right, with the {\em same}
probability $q$ for all incoming channels, while the nodes in
Ref.~\onlinecite{Bocquet} describe scattering to the right with
the probability $q$ from two opposite channels, and with the
probability $(1-q)$ from the other two opposite channels. In terms
of coupling of counter-propagating channels on the links, in Ref.
\onlinecite{Bocquet} this coupling differs from our link matrix in
following respects: to get the coupling of
Ref.~\onlinecite{Bocquet} one should replace in
Eq.~(\ref{Tmatrix}) the phases $\phi_i$ should all be put zero,
while $\varphi_i$ satisfy the relation $\varphi_1=\varphi_2$,
$\varphi_3=\varphi_4$, which is different even from the case with
TRS in the p-q model.

\section{Acknowledgements}
We acknowledge fruitful discussions with I. Gruzberg. This work
was supported by the BSF grant No. 2006201. One of us (V. K.)
appreciates the hospitality of the MPI-PKS in Dresden, where
significant part of his work was done.

\section{Appendix}

The easiest way to derive Eq. (\ref{mfp}) is to notice that with
respect to the motion along the diagonal direction of network the
diffusion has, effectively, a one-dimensional character
\cite{Janssen}. This is because the motion along the diagonal can
be viewed as a random sequence of transmissions and reflections
without deflections. For a given site, the full probability of
transmission along the diagonal direction (say, up and to the
right) is $\mathcal{T}=t^2+d^2$, while the full probability of
reflection is $1-\mathcal{T}=r^2+d^2$. Parameter
$k_{\scriptscriptstyle F}l$ should be identified with ratio,
$\mathcal{T}/(1-\mathcal{T})$. Together with flux-conservation
condition, $t^2+r^2+2d^2=1$, this ratio reduces to
Eq.~(\ref{mfp}).

As we pointed out in Section II, the point $d=1/\sqrt{2}$ is
singular: strong localization predicted by scaling theory with
Boltzmann result Eq. (\ref{mfp}) as  an initial condition,
contradicts to the result of Ref. \onlinecite{Bocquet}. This
contradiction can be resolved from the following reasoning.

The same {\it average} $\langle d^2\rangle$ can be realized in two
completely different ways:

({\it i}) Boltzmann approach. In the two instances of scattering
at the {\it same node}, electron can be deflected both to the left
and to the right with equal probabilities, $50\%$.

({\it ii}) Percolative approach. A given node scatters {\it always
to the left or always to the right}; $50\%$ of the nodes scatter
to the left, and remaining $50\%$ scatter to the right.

Note, that the classical motion of electron is very different in
the above two realizations. In the case ({\it i}) this motion is
diffusive. Then interference will readily localize this motion.
The case ({\it ii}) corresponds to classical percolation
threshold. To see this, notice, that the electron predominantly
travels in loops (clusters) and one trajectory is the infinite
cluster spanning the entire system. In fact, it can be easily seen
that possible trajectories belong to two decoupled
bond-percolation networks. Interference will transform these
networks to two CC models at the delocalization-transition points.
Since the percolative picture is fully coherent, it captures
properly the quantum delocalization, while "dephased" Boltzmann
picture does not.

Thus the distinguished characteristics of the point $d=1/\sqrt{2}$
is that only at this point there is sensitivity to the realization
of the disorder.

Note that in the case $t=r=0$ but $d_1=\sqrt{1- d_2^2}$ not equal
to $1/\sqrt{2}$, prediction of the scaling theory that the
Boltzmann result governs localization properties at large
distances is correct. Although with a strong spread in local
$d$-values, when the network still breaks into two CC networks,
the states of both CC networks are localized with localization
radius, $\xi\sim |d_1^2-1/2|^{-4/3}$.

\end{document}